\documentclass[10pt,aps,prb,twocolumn,superscriptaddress,floatfix]{revtex4-2}
\usepackage{amssymb,amsmath,bm} 
\usepackage{graphicx} 
\usepackage{braket}
\usepackage{hyperref} 
\hypersetup{colorlinks=true,    linkcolor=blue,    citecolor=blue,     urlcolor=blue}
\newcommand{\mk}[1]{\mathbf{#1}}
\begin{document}
 
\title{Multimagnon bound states in the polarized phase of anisotropic antiferromagnets}

\author{A. El Mendili}
\affiliation{Universit\'e Grenoble Alpes, CEA, IRIG, PHELIQS, 38000 Grenoble, France}
\author{M. E. Zhitomirsky}
\affiliation{Universit\'e Grenoble Alpes, CEA, IRIG, PHELIQS, 38000 Grenoble, France}
\affiliation{Institut Laue Langevin, 38042 Grenoble Cedex 9, France}

\begin{abstract}
We investigate single-ion bound magnon states in the high-field polarized phase of uniaxial 
antiferromagnets. By solving the two-particle problem exactly, we determine
the dispersion of bound magnon pairs both below and above the continuum.
We also obtain rigorous bounds on their binding energies by using the Gershgorin circle theorem.
For spin-1 easy-axis systems, two-magnon bound states with zero total momentum emerge as low-energy 
excitations once the ratio of single-ion anisotropy to exchange exceeds a threshold value. 
As the field decreases, magnon pairs condense, driving a transition into a spin-nematic 
phase. Beyond the two-magnon problem, we develop a general strong-coupling approach that describes 
$n$-magnon single-ion bound states ($n\leq 2S$) in large-$S$, easy-axis antiferromagnets.
The Bose-Einstein condensation of these multi-magnon excitations  yields a rich hierarchy of field-induced
multipolar phases stabilized by single-ion anisotropy. 
Comparisons with recent experiments on Na$_2$BaNi(PO$_4$)$_2$ and Sr$_2$CoGe$_2$O$_7$ illustrate 
the quantitative power of our theory.
\end{abstract}
\date{November 5, 2026}
\maketitle

\section{Introduction}

Magnons, the elementary excitations in magnetic solids, can bind together to form new 
composite quasiparticles. These emergent excitations resemble Cooper pairs in superconductors and 
hadrons in high-energy physics, all of which arise as bound states of more fundamental constituents.
Composite magnetic quasiparticles exhibit distinct spectroscopic signatures, appearing as discrete modes outside 
the multimagnon continua. When located in the low-energy part of the spectrum, they can further condense 
upon variations of the applied magnetic field or pressure, producing novel states characterized by 
quadrupolar or higher-order multipolar order parameters rather than conventional dipolar ordering.

Several distinct mechanisms of magnon attraction have been identified to date. 
Exchange bound states (EBS) may arise when two spin flips move to adjacent sites in order to lower
their exchange energy.
Historically, this mechanism was primarily studied for Heisenberg ferromagnets  
\cite{Bethe31,Wortis63,Hanus63,Fukuda63,Torrance69,Wada75,Wada76,Nayak22}
and is included in standard textbooks \cite{Mattis,Rastelli}. Nonetheless, exchange-mediated magnon attraction 
can also manifest in antiferromagnets---both in zero \cite{Oguchi71,Oguchi73,Hamer09} and in 
high magnetic fields \cite{Ono71,Chubukov91,Shannon06,Kecke07,Zhito10,Arlego11,Jiang23,Li25,Gallegos26}. 
Alternatively, in anisotropic magnetic materials  with large spins $S>1/2$,
two or several magnons  can be attracted to the same site by the single-ion term,
leading to single-ion bound states (SIBS) 
\cite{Silberglitt70,Tonegawa71,Papan87,Bahurmuz87,Wierschem12,Sharma22,Anas25,Wang26}. 
Once the easy-axis anisotropy and the exchange constant are comparable, $D\simeq J$, an additional 
branch of multipolar excitations or longitudinal magnons can emerge
 \cite{Wyzula22,Anas25}. A longitudinal magnon can be visualized as a full spin reversal, $S \to -S$,
propagating across the lattice. Equivalently, it can be represented as a composite bound state of $2S$ single
magnons.

In the presence of repulsive magnon interactions, bound states 
can appear in the high-energy part of the spectrum, above the two-magnon continuum \cite{Papan87,Papan89,Halati23}. 
These high-energy excitations are often called repulsively bound magnon states.
Finally, the correlated pair hopping mechanism  is operative for certain frustrated spin dimer models producing
bound states of two triplons ($S=1$ magnons)  \cite{Knetter00,Totsuka01}.

Over the years, bound states of magnons have been discovered in a variety of magnetic materials.
Multimagnon EBS have been observed  in 
CoCl$_2\cdot 2$H$_2$O \cite{Torrance69b}, 
$\alpha$-NaMnO$_2$  \cite{Dally20},  $\alpha$-RuCl$_3$ 
\cite{Saha20}, and CoV$_3$O$_8$ \cite{Xiao26}. Two-magnon SIBS have been 
found in the effective $S=1$ materials FeBr$_2$ \cite{Psaltakis84} and  FeI$_2$ \cite{Bai21}, 
while the longitudinal magnons (4-magnon SIBS) have been observed in the $S=2$ antiferromagnets 
FePS$_3$ \cite{Wyzula22},  FePSe$_3$ \cite{LeMardele24}, and LiFePO$_4$ \cite{Peedu22}.
Repulsively bound magnons have been identified in the ESR spectra of several
nickel  \cite{Orendac99,Zvyagin07} and cobalt materials
\cite{Akaki17,Wang24}.
Lastly, kinetically bound magnons exist in the Shastry-Sutherland lattice material 
SrCu$_2$(BO$_3$)$_2$ \cite{McClarty17}.  Furthermore, the direct realization of bound magnons in optical 
experiments with cold atomic gases \cite{Fukuhara13,Kranzl23}
has renewed general interest  in multimagnon excitations in quantum spin models
\cite{Qin17,Dong23,Mook23,Wu22,Li24,Lou24}.

Of particular significance is the recently observed  Bose-Einstein condensation (BEC) of bound magnon pairs
in a magnetic field  in 
the spin-1 triangular-lattice antiferromagnet Na$_2$BaNi(PO$_4$)$_2$ \cite{Sheng25a,Sheng25b}. 
Below the condensation field, a quantum spin-nematic phase is formed, providing a remarkable 
example of unconventional quadrupolar order in magnetic systems \cite{Blume69,Andreev84,Penc11}. 
This discovery raises an important question: what physical conditions determine 
the emergence of the spin nematic phase or other multipolar states in anisotropic antiferromagnetic materials?
To address this question, we develop a unified theoretical description of multimagnon bound states 
in the high-field polarized phase 
of easy-axis antiferromagnets with large spins $S\geq 1$.

To this end, we study the excitation spectrum of an antiferromagnet described by the anisotropic spin Hamiltonian:
\begin{eqnarray}
\hat{\mathcal{H}} & = & J \sum_{\langle ij\rangle}\bigl[ S^z_i S^z_j + \lambda(S^x_i S^x_j + S^y_iS^y_j)\bigr] 
- D\sum_i({S}_i^z)^2
\nonumber\\
    & &\qquad \mbox{}  - H \sum_i S_i^z\ ,
\label{H0}    
\end{eqnarray}
where ${\bf S}_i$ are spin-$S$ operators and ${\langle ij\rangle}$ denotes summation over nearest-neighbor 
bonds on a Bravais lattice with $N$ sites. Our main focus is on the easy-axis anisotropy: $D>0$ and 
$\lambda\leq 1$.  We also present new results for the easy-plane case, $D<0$,  and for ferromagnets, $J<0$.

\begin{figure}[t]
\centerline{
\includegraphics[width=0.45\columnwidth]{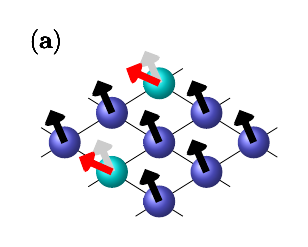}
\hskip 1mm
\includegraphics[width=0.45\columnwidth]{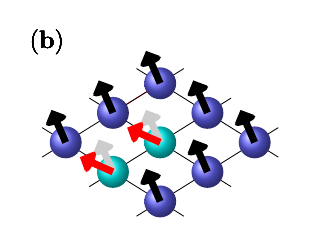}
}\centerline{  
\includegraphics[width=0.45\columnwidth]{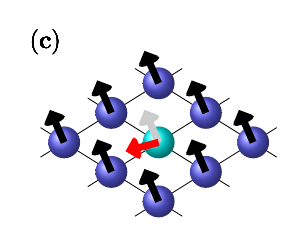}
\hskip 1mm
\includegraphics[width=0.45\columnwidth]{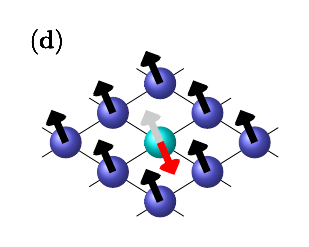}
}\caption{Schematic representation of multi-magnon states in the uniformly polarized phase: 
(a) two independent magnons; (b) a pair of magnons in the exchange bound state; (c) a pair of magnons 
in the single-ion bound state; (d) a longitudinal magnon equivalent to the single-ion bound state of 
$2S$ magnons.
}
\label{fig_cartoon}
\end{figure}

A simple, qualitative picture of the possible multi-magnon excitations is constructed by starting with 
the Ising limit in (\ref{H0}).  Due to the spin rotation symmetry about the $\hat{z}$-axis, all eigenstates 
can be assigned with a definite $S^z$ value. In the Ising approximation, one magnon or a single spin flip 
with $S^z=1$ ($S^z_{\rm tot} = NS -1$) has the excitation energy 
\begin{equation}
E_1=H + D(2S-1) - zJS \ ,
\label{E1}
\end{equation}
where $z$ is the lattice coordination number. The magnon kinetic energy is determined by
the transverse terms and is small for $\lambda\to 0$. Two independent spin flips in the $S^z=2$
subsector, Fig.~\ref{fig_cartoon}(a), form a two-particle continuum centered around 
$E_2 = 2E_1$ with a width $O(\lambda J)$. For the nearest-neighbor antiferromagnet ($J>0$),
two adjacent spin flips, Fig.~\ref{fig_cartoon}(b), have an increased energy $E_{2a} = 2E_1 + J$.
For small $\lambda$, their energy $E_{2a}$ lies outside of the two-magnon continuum. 
The corresponding states form a high-energy branch of repulsively bound magnons. 

Similarly, two spin flips on the same site, Fig.~\ref{fig_cartoon}(c), lower their energy 
from $2E_1$ to
\begin{equation}\label{E_2b}
E_{2b} = 2E_1-2D \ .
\end{equation}
Accordingly, the single-ion anisotropy of the easy-axis type provides an attractive potential for magnon pairs. 
For small enough magnon hopping $\lambda J$, this leads to an extra branch of bound states below 
the two-magnon continuum.

The single-ion term produces attraction also in all other $S^z$-subsectors. For $n$ magnons,
$2\le n\le 2S$, a state with all spin flips on the same site has energy
\begin{equation}
E_{nb} = nE_1-n(n-1)D \ .
\label{En}
\end{equation}
Hence, a branch of  bound states can be formed in each of the $n$-magnon subsectors. 
The binding energy  (\ref{En}) is largest for $n=2S$, which is equivalent to the complete spin reversal 
$\ket{S}\rightarrow\ket{-S}$, see Fig.~\ref{fig_cartoon}(d). These excitations called longitudinal
 ($L$) magnons develop into the lowest energy branch for $D\agt J$ \cite{Anas25}.

In contrast to the weak-coupling problem \cite{LL3}, where the existence of bound states is strongly affected 
by spatial dimensionality, the strong-coupling regime of the spin model (\ref{H0}) is governed primarily by the lattice 
topology.  In the Ising limit  $\lambda\ll 1$, the two-magnon sector contains one SIBS branch and $z/2$ EBS 
branches. The latter number is fixed by the number of independent nearest-neighbor bonds rather than by 
the lattice dimensionality, see Sec.~\ref{sec:2mag}. This simple bond rule holds up to a certain value $\lambda_c\alt 1$, 
above which some of the bound-state branches start to merge with the continuum. At the Heisenberg point $\lambda=1$, 
magnons acquire substantial kinetic energy and SIBS can appear only above a threshold value $(D/J)_c$.
Below, we determine the corresponding threshold values for the spin-$1$ Heisenberg chain and for 
spin-$1$ antiferromagnets on square and triangular lattices. We also develop a strong-coupling expansion 
that extends the analysis to models with larger spin values and to bound states of more than two magnons. 
Studying the stability of these multi-magnon excitations helps identify the range of microscopic parameters for 
which unconventional multipolar phases can appear in strong magnetic fields.

The remainder of the paper is organized as follows. In Sec.~\ref{sec:2mag}, we revisit the analytical solution of 
the two-magnon problem and, in particular, derive new rigorous bounds on the energies of magnon pairs, 
applicable to both ferromagnets and antiferromagnets. In Sec.~\ref{sec:Multi} we develop a strong-coupling expansion for 
the excitation energies of
$n$-magnon single-ion bound states located both below ($D>0$) and above ($D<0$) the two-magnon continuum. In Sec.~\ref{sec:Num} 
numerical solutions of the exact two-magnon equations are compared with the approximate results for 
spin-$1$, $3/2$, and $2$ models. We further determine the conditions under which multimagnon bound states become the leading instability of the polarized phase. Their condensation gives rise to a spin-nematic phase for $S=1$ and to higher multipolar phases for $S>1$. A brief discussion of two experimental examples of bound magnon
states in  Na$_2$BaNi(PO$_4$)$_2$ and Sr$_2$CoGe$_2$O$_7$ is included in Sec.~\ref{sec:Exp}. 
The final conclusions are presented in Sec.~\ref{sec:Sum}.

\section{Two-magnon bound states}
\label{sec:2mag}
\subsection{Exact solution}
\label{sec:ES}

The bound-state problem can be solved exactly  for two magnons in the fully polarized phase of a spin-$S$ uniaxial
antiferromagnet  (\ref{H0}). We denote this quantum state as  $\ket{0} = \ket{\uparrow\uparrow\uparrow\ldots}$. Its energy 
coincides with the classical value 
\begin{equation}
  E_0/N = zJS^2/2 - D S^2 - HS \ ,
\label{Egs}    
\end{equation}
where $N$ is the total number of spins. The eigenstates in the one-magnon ($S^z =1$) subsector are
plane waves
$\ket{1_{\bf k}}\sim \sum_i e^{-i{\bf k}{\bf r}_i} S_i^-\ket{0}$ with
the excitation energy
\begin{equation}
\epsilon_{\mk{k}} = \varepsilon_0 + z\lambda JS\,\gamma_{\mk{k}} \ .
\label{TM1}
\end{equation}
Here $\varepsilon_0 = H + D(2S-1) -  zJS$ is the Ising energy of one spin deviation, cf.\ Eq.~(\ref{E1}),
and the lattice function $\gamma_{\mk{k}}$ is defined by
\begin{equation}
\gamma_{\mk{k}} = \frac{1}{z} \sum_{\bm{\rho}} e^{i\mk{k}\bm{\rho}} 
\end{equation}
with summation performed over $z$ nearest-neighbor bonds denoted by $\bm{\rho}$.  The magnon energy 
vanishes at the critical field
\begin{equation}
H_{c1} = zJS (1-\lambda\gamma_{\mk{Q}}) - D(2S-1) \,,
 \label{Hc1}
\end{equation}
where  $\gamma_{\mk{Q}} = \min_{\mk{k}}\{\gamma_{\mk{k}}\}<0$. 
Magnons  with the momenta ${\bf k} = \mk{Q}$ undergo 
Bose-Einstein condensation in this field,  transforming the fully polarized phase 
into a canted  antiferromagnetic state  \cite{Zapf14}. However, for large $D/J$, the transition at $H_{c1}$
may be superseded  by condensation of multimagnon excitations
opening a route to exotic quantum ground states.

A general two-magnon state can be written  as 
\begin{equation}
\ket{2}=\sum_{i,j} \psi_{ij}\,{S}^-_i{S}^-_j\ket{0}  =
\sum_{i,j} \psi_{ij}\ket{2_{ij}} \,,
\label{2mag} 
\end{equation}
where the double summation over the site index is assumed. 
Since spin operators on different sites commute, the amplitudes must  satisfy $\psi_{ij}=\psi_{ji}$. 
The states $\ket{2_{ij}}$ form an orthogonal basis in the two-magnon subsector, which has a total dimension of $N(N+1)/2$.
We use these states without normalization factors, which simplifies the equations on $\psi_{ij}$ below.
The eigenstate condition yields
\begin{equation}
E  \ket{2}=\sum_{i,j}\psi_{ij}[\hat{\mathcal{H}},{S}^-_i{S}^-_j]\ket{0} \,,
 \label{Eigen}
\end{equation} 
where  $E$ is the excitation energy measured relative to the ground state (\ref{Egs}).
Computing  commutators and multiplying both sides of (\ref{Eigen}) with $\bra{2_{ij}}$,
one obtains a set of difference equations on $\psi_{ij}$.
These equations are further transformed by assigning a total momentum $\mk{k}$ 
to the two-magnon state:
\begin{equation}
\psi_{ij} = \frac{1}{\sqrt{N}}\, e^{i\mk{k}\,(\mk{r}_i+\mk{r}_j)/2}\,  \psi_{\bf k} ({\bf r}) \ ,
\label{FTK}
\end{equation} 
where  $\mk{r}=\mk{r}_j-\mk{r}_i$ is a relative distance between two spin deviations. 
The site permutation symmetry implies $\psi_{\bf k} (-{\bf r}) =  \psi_{\bf k} ({\bf r})$. 
The two-magnon Schr\"odinger equation is then expressed as 
\begin{eqnarray}
&& (E - 2 \varepsilon_0) \psi_{\bf k} ({\bf r}) = \lambda JS 
\sum_{\bm{\rho}} c_{\bm{\rho}}\bigl[  \psi_{\bf k}({\bf r}-\bm{\rho}) + \psi_{\bf k}({\bf r}+\bm{\rho})\bigr] 
\nonumber
 \\
& & \mbox{} + J
  \sum_{\bm{\rho}} \delta_{{\bf r},\bm{\rho}}  \bigl[\psi_{\bf k} (\bm{\rho}) - \lambda 
  c_{\bm{\rho}}\psi_{\bf k} (0)\bigr] - 2D\psi_{\bf k} (0)  \delta_{{\bf r},0}\,,
\label{E2Kr} 
\end{eqnarray}
where the short-hand notation  $c_{\bm{\rho}}  = \cos ({\bf k}\!\cdot\!\bm{\rho}/2)$
is used and $\delta_{{\bf x},{\bf y}}$ denotes the Kronecker symbol.
The first term on the right-hand side 
represents the kinetic energy of magnons,
whereas the last two terms describe the magnon interactions.
Equation (\ref{E2Kr})  is valid for either sign of the microscopic  constants $J$ and $D$ as long as 
the spins are fully polarized by an applied magnetic field.
It can therefore describe both attractively and repulsively bound states.
The magnitude of the magnon interaction is reduced by a factor of $1/S$ relative to their kinetic energy.
Consequently, bound states become more fragile as the spin value $S$ increases.
Likewise, the kinetic energy contains the lattice coordination factor $z$, which is absent in the
interaction terms. Thus, bound state formation is generally suppressed in higher dimensions, 
although frustration effects may modify a simple monotonic dependence on  $z$.

Fourier transformation with respect to the relative coordinate $\bf r$  
\begin{equation}
\psi_{\bf k} ({\bf r}) = \frac{1}{N} \sum_{\bf q}  e^{i\mk{q}\mk{r}}\,\psi_{\mk{k},\mk{q}}  
\label{FTp}
\end{equation}
converts (\ref{E2Kr}) into 
 \begin{eqnarray}
&&
\bigl(E\! -\epsilon_{\mk{k}/2+\mk{q}}\! - \epsilon_{\mk{k}/2-\mk{q}}\bigr) \psi_{\mk{k},\mk{q}}
=  \frac{1}{2N}\! \sum_{\mk{p}}  \Bigl[ z J(\gamma_\mk{p+q}\!   + \gamma_\mk{p-q}) 
\nonumber \\
&& \mbox{\quad\quad}
-z\lambda J
(\gamma_{{\bf k}/2+{\bf q}} + \gamma_{{\bf k}/2-{\bf q}}) - 4D
\Bigr]\psi_{\mk{k},\mk{p}}\,.
 \label{E2Kp}
\end{eqnarray}
This integral equation is a convenient starting point for numerical solution of  the two-magnon bound state problem.
The Fourier amplitudes  $\psi_{\mk{k},\mk{q}}$ are expanded in a sum of few
 terms proportional to lattice harmonics of $\bf q$ appearing on the right-hand-side of Eq.~(\ref{E2Kp}),
 which leads to a secular equation on $E$ 
  \cite{Fukuda63,Mattis,Rastelli}.  

Alternatively, the bound-state energies can be obtained directly from the Schr\"odinger equation in real space 
\cite{Torrance69,Wu22,Li24,Lou24,Anas25}. In this case, the wave function
is constructed on a finite $N$-site lattice with the open boundary conditions.  
The amplitudes  $\psi_{\bf k} ({\bf r})$ are arranged into 
a one-dimensional array of length $N$ and the Schr\"odinger equation (\ref{E2Kr}) is represented as
the eigenvalue problem for a block-tridiagonal $N\times N$ matrix $\mathbf{M}$. 
Among its eigenstates,  only even parity solutions correspond to real physical states. 
An explicit example of matrix $\mathbf{M}$ is presented 
in the next subsection.  Within numerical accuracy, the two approaches produce equivalent results for 
the bound state energies and require comparable amounts of computation time. The bound state solutions obtained 
by numerically solving equations (\ref{E2Kr}) and (\ref{E2Kp}) are described in Sec.~\ref{sec:Num}. The rest of 
this section discusses general properties of these states. 

A natural question that arises in this context is how many  branches of two-magnon bound states
can exist in the two-magnon subsector of a $d$-dimensional spin model. Early theoretical works \cite{Wortis63,Hanus63,Fukuda63,Torrance69} found that nearest-neighbor Heisenberg ferromagnets on 
a linear chain, square lattice, and simple-cubic lattice possess, respectively, one, two, and three distinct  
EBS branches  in the two-magnon subsector. One possible interpretation of these results 
is that the number of bound-state branches is equal to the dimensionality $d$ of a spin model. 
However, there are various counterexamples to this rule in the literature, showing numbers of 
branches that are smaller \cite{Wada75,Wada76} or larger \cite{Bahurmuz87,Li25} than $d$. 
Nevertheless, the dimension rule remains valid for simple spin models on hypercubic lattices. 
Its origin in this case can be traced to a special property of the two-magnon continuum.
At the Brillouin-zone corner ${\bf k}^*=(\pi,\pi,\ldots)$, the continuum shrinks to a single point. 
Then,  the integral equation (\ref{E2Kp}) is greatly simplified providing solutions
according to the number of independent bond directions equal to $d$  \cite{Gallegos26}.

Here, we obtain a general answer to the question about the maximal number of  the bound state solutions 
by analyzing the  Schr\"odinger equation in real space.  In the Ising limit  $\lambda=0$, 
the eigenvalue matrix defined by Eq.~(\ref{E2Kr}) is diagonal. The corresponding eigenstates describe 
two spin deviations occupying either the same site, with $E=2\varepsilon_0-2D$ (SIBS), or adjacent sites connected 
by the exchange bond $\bm{\rho}$, with $E=2\varepsilon_0+J$ (EBS). Depending on the sign of microscopic constants, 
these solutions represent either attractively
bound states with energies below the two-magnon ``continuum'' at $E=2\varepsilon_0$, 
or repulsively bound states  lying above. 
The total number of independent EBS states in the Ising limit is simply equal to the number of exchange bonds, $N_{b} = zN/2$, 
where $z$ is the lattice coordination number.

For small values of $\lambda$, spin deviations delocalize and form a magnon band with a bandwidth proportional 
to $\lambda J$. Since the binding energies in the Ising limit are set by $D$ or $J$, the bound-state branches 
remain separated from the two-magnon continuum in a finite range of $\lambda\leq\lambda_c$. In 
this regime, the number of EBS branches is determined by a simple bond rule: 
The total number of exchange bonds $N_b$ must be divided by the number of allowed momenta in the Brillouin zone. 
For a nearest-neighbor model, $N_b=zN/2$, while for a Bravais lattice the number of momenta equals 
$N$, giving $z/2$ distinct EBS branches. 
Thus, the square-lattice model ($z=4$) has at most two EBS branches, whereas the triangular lattice (anti)ferromagnet
($z=6$) exhibits three branches for $\lambda<\lambda_c$. 
The dimension rule for hypercubic lattices ($z=2d$) therefore appears as a special case of this more
general bond-counting principle. The same reasoning 
can be applied to non-Bravais lattices and to models with several exchange shells. In addition, the single-ion term in 
the Hamiltonian (\ref{H0}) produces an extra SIBS branch, which remains well defined as long as $\lambda J\alt D$.

The peculiar property of the  two-particle excitation spectrum of hypercubic models
near the Brillouin zone corner ${\bf k}^*=(\pi,\pi,\ldots)$ also has a simple interpretation
 in the real-space approach.
The cosine factors $c_{\bm{\rho}}^*  = \cos ({\bf k}^*\!\cdot\bm{\rho}/2)$ vanish simultaneously for all 
nearest-neighbor vectors $\bm{\rho}$, and Eq.~(\ref{E2Kr}) reduces to the diagonal Ising form.
Consequently, the Ising energy spectrum re-emerges close to this point:
the energies of single-ion and exchange bound states are given, respectively, by $-2D$ and $J$.
 By the same token, the two-magnon continuum
\begin{eqnarray}
 E_2(\mk{k},\mk{q}) & =  & \epsilon_{\mk{k}/2+\mk{q}} + \epsilon_{\mk{k}/2-\mk{q}} 
 \nonumber \\[1mm]
 & = &  2\varepsilon_0  + 2 \lambda J S  \sum_{\bm{\rho}} 
c_{\bm{\rho}} \cos (\mk{q}\! \cdot\!\bm{\rho})
\label{E2kq}
\end{eqnarray}
shrinks to a single pinch point at ${\bf k}={\bf k}^*$ exhibiting a bow-tie shaped structure
in the frequency-momentum plane. These pinch points reduce
the integral equation (\ref{E2Kp}) to an algebraic one, producing
the Ising structure of energy levels \cite{Gallegos26}.

The protected Ising spectrum and the continuum pinch points are, however, specific to the  spin models 
on hypercubic lattices. For instance, the triangular lattice has three different bond orientations and
no $(k_x,k_y)$ exists, for which $c_{\bm{\rho}} \equiv 0$ simultaneously for all six
nearest-neighbor vectors $\bm{\rho}$. 
Therefore, the Ising energy spectrum cannot reappear for any  ${\bf k}$ and
some of the bound-state branches may disappear completely once $\lambda$ increases from 0
to 1. The disappearance of the bound state branches depends on the position and shape of the continuum boundaries. 
Consequently, this behavior is model-dependent and may proceed differently for bound magnons above and 
below the continuum.

\subsection{Exact energy bounds}
\label{sec:EEB}

We now use the matrix form of the  Schr\"odinger equation to derive exact bounds on the energies of the single-ion and
exchange bound states. To this end,  let us recast the difference equations  as an eigenvalue
problem 
\begin{equation}
E \Psi_{\bf k} = \mathbf{M}\,\Psi_{\bf k}  \ .
\end{equation}
We will illustrate this construction for a spin chain.
Consider $N$ spins  on a line with open boundary conditions, and assign the site coordinates 
as follows
\begin{equation}
r = \bigl\{-(N-1)/2,   \ldots,  0,\ldots, (N-1)/2\bigr\},
\label{rL} 
\end{equation}
with $N$ assumed odd, for convenience. The amplitudes are arranged 
in the column vector $\Psi_{\bf k} = \{\psi_{\bf k}(r)\}$.
The difference equations  (\ref{E2Kr}) can then be represented by the  tridiagonal $N\times N$  matrix:
\begin{equation}
\mathbf{M} =
\begin{pmatrix}
A & B &           &           &          &        &          &       &      \\
B & A &    B     &           &          &        &          &       &      \\
   &    & \ddots &           &          &        &          &       &      \\
   &    &     B     &  A_1  &   B_1 &        &          &       &      \\
   &    &            &   B      & A_0 &  B    &           &       &      \\
   &    &            &           &  B_1 & A_1 &   B      &      &     \\
   &    &            &           &          &        &\ddots &      &      \\
   &    &            &           &          &        &   B     &   A  &  B \\
   &    &            &           &          &        &           &  B  &  A \\
\end{pmatrix}\!,
\label{MN}
\end{equation}
where only the nonzero elements  are displayed: 
$A=2\varepsilon_0$, $A_0=2\varepsilon_0-2D$,  $A_1=2\varepsilon_0+J$,  
$B  =  2\lambda JS \cos(k/2)$, $B_1 = B[1-1/(2S)]$,
and diagonal dots stand for repeating tridiagonal rows $(B,A,B)$.

The square matrix $\mathbf{M}$ belongs to the class of centro\-symmetric matrices that satisfy
\begin{equation}
M_{ij}=M_{N+1-i,N+1-j} 
\end{equation}
or, equivalently, remain invariant under inversion of all matrix elements with respect to the central element $M_{(N+1)/2,(N+1)/2}$
[$A_0$ for Eq.~(\ref{MN})]. 
The eigenvectors of a centrosymmetric matrix  can be chosen to have definite parity under this inversion
\cite{Andrew73}.
For odd $N$,  there are $(N+1)/2$ symmetric (even) eigenvectors and $(N-1)/2$ skew-symmetric (odd) eigenvectors. 
Only the even solutions describe the physical two-magnon states, which possess the required symmetry of the wave function under 
permutation of the two sites.  Since there are $N$ distinct values of the total momentum $\bf k$,
 the total number of physical states is $N\times(N+1)/2$, in agreement with the dimension of the real-space basis.

The above construction can be readily generalized to two dimensional models. For the square-lattice antiferromagnet, 
the column vector $\Psi_{\bf k}$ is constructed by sequentially rearranging the rows of an $L\times L$ cluster into 
a one-dimensional array. Then, the matrix ${\bf M}$  acquires the block tridiagonal form analogous to Eq.~(\ref{MN}), 
with each scalar entry replaced by 
a $L\times L$ block. For example, the diagonal `bulk' block $\bm{A}$ is itself tridiagonal
 \begin{equation}
A_{ij} = 2\varepsilon_0 \delta_{i,j} + 2\lambda JSc_x (\delta_{i,j+1} + \delta_{i,j-1}) \ , 
\end{equation}
whereas  the adjacent block is $\bm{B} = 2\lambda JSc_y \bm{1}$ with  $c_{x,y} = \cos(k_{x,y}/2)$.
The matrix elements of extra blocks $\bm{A}_{0}$, $\bm{A}_{1}$, and $\bm{B}_1$ 
associated with two spin flips on the same or the nearest-neighbor sites have to be modified accordingly. 
Generally, they  preserve the centrosymmetric structure of the full matrix  ${\bf M}$.
Their explicit form is not needed below and can be easily obtained following  Ref.~\cite{Anas25}.

To obtain rigorous bounds on the two-magnon energies, we employ the Gershgorin circle theorem \cite{Gershgorin31,Varga04},
which states that any eigenvalue of a general matrix $\mathbf{M}$ is contained in one of the intervals centered at diagonal elements:
\begin{equation}
|E - M_{ii}| \leq \sum_{j\neq i}|M_{ij}| \ .
\label{Gershgorin}
\end{equation}
Thus, without solving the full eigenvalue problem, one can constrain the energies by inspecting matrix elements row by row. 
In the above spin-chain example (\ref{MN}), the $(B,A,B)$ rows constrain the energies of 
the propagating two-magnon states and, hence, block the boundaries of the two-magnon continuum. 
The central row with the shifted diagonal element $A_0$, puts a restriction on the energy of the single-ion bound state. 
Likewise, the two rows containing $A_1$ determine the energy interval relevant to the exchange bound states. 

After illustrating the procedure with a simple one-dimensional example, we turn to a general  spin model. 
The Gershgorin bounds can be formulated directly from the difference equations  (\ref{E2Kr}) without explicitly 
constructing the corresponding matrix $\mathbf{M}$. For any value of the total momentum $\bf k$, 
the energy of the two-magnon bound state must belong to one of the two intervals
\begin{equation}
|E - 2 \varepsilon_0 + 2D|   \leq   2\lambda S  \sum_{\bm{\rho}}  | J c_{\bm{\rho}}| 
\label{Gershgorin1} 
\end{equation}
and
\begin{equation}
|E - 2 \varepsilon_0 - J| \leq  2\lambda S   \Bigl(\sum_{\bm{\rho}} | J c_{\bm{\rho}}|\Bigr)   
- \lambda |J c_{\bm{\rho}'}|
\label{Gershgorin2} 
\end{equation}
where  ${\bm{\rho}'}$ is one of $z$ nearest-neighbor vectors. 
The bounds (\ref{Gershgorin1}) and (\ref{Gershgorin2}) become particularly useful 
whenever one of the boundaries of the Gershgorin interval for the bulk states coincides with 
the  lower $E_2^l({\bf k})$ or upper $E_2^u({\bf k})$ boundary of the two-magnon 
continuum $E_2({\bf k},{\bf q})$. 
In such cases, the Gershgorin bounds directly restrict the possible values of 
the binding energy. Two specific examples are provided below.

\subsubsection{XXZ ferromagnet}

We begin with the nearest-neighbor $XXZ$ ferromagnet (\ref{H0})  on a Bravais lattice  with  $J<0$ and $D=0$. 
Useful physical results can be derived from Eqs.~(\ref{Gershgorin1}) and (\ref{Gershgorin2}) when the
 total momentum  is zero ${\bf k}=0$. In this case, the cosine factors are trivial $c_{\bm{\rho}} = 1$ and the bottom of the two-magnon 
continuum (\ref{E2kq}) is attained for ${\bf q}=0$:
\begin{equation}
E_{2}^l(0)=2\varepsilon_0-2z\lambda |J|S\,.
\end{equation}
Importantly, this energy coincides exactly with the lower boundary of the Gershgorin interval generated by the ``bulk" rows of matrix 
$\bf M$. Hence, the lower boundary of the interval (\ref{Gershgorin2}) for an exchange bound state can be compared directly with the actual continuum edge. Introducing a non-negative binding energy
\begin{equation}
E_B = E_{2}^l(0) - E\,,
\end{equation}
one obtains
\begin{equation}
E_B \leq |J|(1-\lambda)\,.
\label{EB_FM}
\end{equation}
Thus, the binding energy at ${\bf k}=0$ is progressively suppressed by the magnon kinetic energy and vanishes in the Heisenberg limit $\lambda=1$.
In particular, the Gershgorin bound (\ref{EB_FM}) proves that 
 a nearest-neighbor Heisenberg ferromagnet has no exchange bound state at zero total momentum.
This result, of course, agrees with the numerical solutions  of the exact equations, 
which find $E_B({\bf k})\to 0$ as $|{\bf k}|\to 0$ for spin chain and 
square-lattice models  \cite{Wortis63,Hanus63,Fukuda63} and no bound states 
 at small $k$ in Heisenberg ferromagnets  on cubic 
\cite{Wortis63} and 
triangular \cite{Wada75} lattices.
 
The same analysis can also be performed for finite $D>0$. 
The relevant Gershgorin interval for SIBS is given by 
(\ref{Gershgorin1}).
Its lower bound coincides with the bottom of the continuum and, consequently,
\begin{equation}
E_B \leq 2D\,.
\label{EB_SI}
\end{equation}
Unlike the EBS case, the Gershgorin bound  puts no constraint that the SIBS has to merge with the continuum as 
$\lambda\to1$. At the same time, this inequality does not imply that $E_B$  remains finite for any value of $D$.

Both relations, (\ref{EB_FM}) and (\ref{EB_SI}), are independent of the lattice and of the spin value $S$.  
By straightforward generalization, the same bounds  also apply to repulsively bound states in the fully polarized phase of easy-plane antiferromagnets, obtained for $J>0$ and $D<0$. In this case, the binding energy is measured from the upper boundary of the continuum
$E_B= E- E_{2}^u(0)$.

\subsubsection{Hypercubic lattices}

For a $d$-dimensional hypercubic lattice, the  above analysis  
 can be extended to arbitrary total momentum $\bf k$. 
 The cosine factors $c_{\bm{\rho}}$ are nonnegative throughout the first Brillouin zone. 
The lower (upper) boundary of the two-magnon continuum is attained once all 
$q$-components in (\ref{E2kq}) are taken simultaneously 0 or $\pi$, depending on the sign of $J$. 
Consequently, both boundaries 
of the ``bulk'' Gershgorin interval coincide with the actual boundaries of the two-magnon continuum.
This immediately implies that the binding energy of SIBS is constrained by 
\begin{equation}
E_B \leq 2|D|  
\label{EBh1}
\end{equation}
 for arbitrary  $\bf k$ and for both 
attractively and repulsively bound states.
In the case of  EBS, one obtains 
\begin{equation}
E_B \leq |J| \bigl(1 - \lambda \min_{\bm{\rho}} \{c_{\bm{\rho}}\}\bigr) .
\label{EBh2}
\end{equation}
At ${\bf k}=0$, this relation reduces to Eq.~(\ref{EB_FM}).  The two bounds 
(\ref{EBh1}) and (\ref{EBh2})
become active (saturated)
at the Brillouin zone boundary for ${\bf k}={\bf k}^*$.

\section{Multi-magnon bound states: \\[1mm] perturbation theory}
\label{sec:Multi}

In this section, we use perturbation theory in the exchange anisotropy parameter $\lambda$
to compute the dispersion of $n$-magnon single-ion bound states for
$n=2,\ldots,2S$ and arbitrary spin value $S$.
We split the Hamiltonian (\ref{H0}) into the Ising part, diagonal in the local $S_i^z$ basis,
\begin{equation}
\hat{\mathcal{H}}_0 =  J\sum_{\langle ij\rangle} S_i^zS_j^z  - D \sum_i(S_i^z)^2  - H \sum_i S_i^z\ ,
\label{H0_nSIBS}
\end{equation}
and the transverse part
\begin{equation}
\hat{V} = \frac{\lambda J}{2} \sum_{\langle ij\rangle} \bigl( S_i^+S_j^- + S_i^-S_j^+ \bigr)
\label{V}
\end{equation}
treated as perturbation. 

The unperturbed Ising energy of a local spin
deviation $\ket{S}\to\ket{S-n}$, which represents a $n$-magnon SIBS,  is given by
\begin{equation}
\varepsilon_{n,0} = nH + nD(2S-n) - nzJS \  .
\label{En0}
\end{equation}
There are $N$ degenerate states of this type on the lattice, corresponding to placing the local
$n$-magnon flip on an arbitrary site. The transverse exchange interactions
generate an effective tight-binding dynamics within this manifold transforming it into a dispersive band.

We retain only the leading nonvanishing contributions to the effective tight-binding Hamiltonian. The local 
bond energy is
renormalized in second order in $\lambda$, whereas the transfer of the entire $n$-magnon flip
between adjacent sites first appears in the $n$th order. To these orders, the corresponding
processes depend only on a single exchange bond and therefore have a general form independent of the
specific lattice geometry. Lattice-dependent processes involving several bonds arise in higher orders
and can be incorporated systematically, if necessary.

As will be made explicit below, the actual expansion parameter is
\begin{equation}
\alpha = \frac{\lambda J}{4D+2J} \  ,
\label{alpha}
\end{equation}
where the denominator is determined by the energy level spacing.
Effectively, we compute the leading term of a general linked cluster series \cite{Oitmaa}.
The present approach should thus be distinguished from the expansion in $J/D$,
which is commonly used to derive an effective $s=1/2$ model for large-$S$ antiferromagnets with strong easy-axis
anisotropy \cite{Wierschem12,Anas25,Damle06}.
The $J/D$ expansion targets the lowest-energy states across all total-spin
subsectors $0\leq S^z_{\rm tot}\leq NS$, whereas we construct an effective tight-binding description
of the $n$-magnon SIBS separately in each subsector.  Although less general in this respect, the present
expansion can remain quantitatively useful even for $D\sim J$ and $\lambda\sim1$, provided that
$|\alpha|$ is still sufficiently small. This feature makes it particularly suitable for comparison with experimental systems.

\subsection{Low-energy expansion}
\label{sec:LowE}

For strong easy-axis anisotropy $D/J> 1$, the $n$-magnon SIBS
are the lowest energy excitations in each total $S^z$ subsector. 
For one bond, the second-order correction to the local energy is generated by
the virtual excitation
\begin{equation}
\ket{(S-n),S} \leftrightarrow  \ket{(S-n+1),(S-1)}
\end{equation}
whereas the hopping amplitude of the $n$-magnon state is determined by the $n$th-order process
\begin{equation}
\ket{(S-n),S}  \to \ldots \to \ket{S,(S-n)}  \ .
\end{equation}
The obtained excitation energy can be represented as 
\begin{equation}
\epsilon_{n}(\mk{k}) = \varepsilon_{n,0}  - nS \sum_{\bm{\rho}} \bigl(t_n e^{i\mk{k}\bm{\rho}} + a_n\bigr).
\label{EnK}
\end{equation}
Here, 
\begin{equation}
\label{an}
a_n =  \frac{(\lambda J)^2}{4D+2J}\,\frac{2S-n+1}{n-1}\ ,
\end{equation}
and
\begin{equation}
\label{tn}
    t_n = \frac{(-\lambda J)^n}{(4D+2J)^{n-1}}\, \frac{(2S-1)!}{(n-1)!(2S-n)!}\ .
\end{equation}

\begin{figure}[tb]
\centering\includegraphics[width=0.95\linewidth]{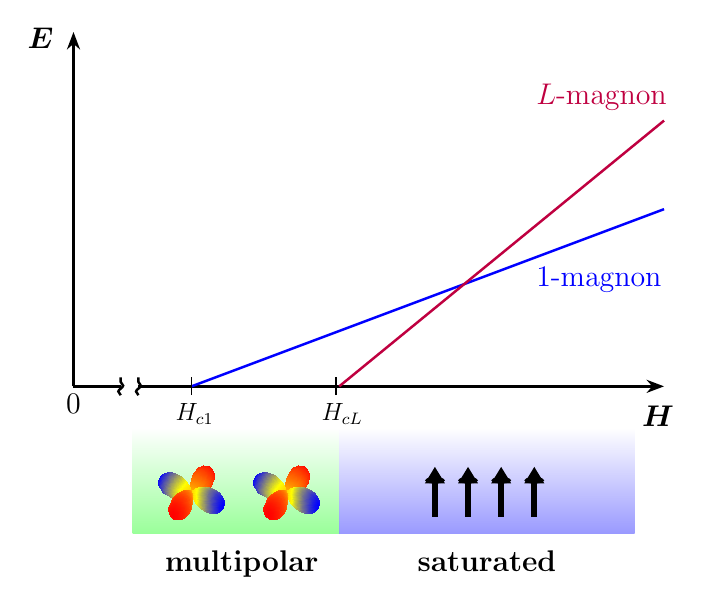}
\caption{Energy-field diagram for Bose-Einstein condensation of $L$-magnons in the
high-field phase of $S\ge 1$ antiferromagnets.
}
\label{fig_BEC}
\end{figure}

An interesting aspect of the above expression is the sign alternation
of the hopping amplitude $t_n$  depending on the parity of  $n$. For even $n$, 
the minimum of the $n$-magnon SIBS branch $\epsilon_{n}(\mk{k})$ is located at ${\bf k}_n=0$, whereas for odd $n$ the minimum
appears at   ${\bf k}_n= {\bf Q}$, same as for the one-magnon branch, Eqs.~(\ref{TM1}) and (\ref{Hc1}).
We further define the excitation gap $\Delta_n =  \epsilon_{n}({\bf k}_n)$, which can, in principle,
be measured in experiments. As follows from Eq.~(\ref{En0}),
the gap  has a linear dependence on a magnetic field and can be written as
\begin{equation}
\Delta_n =  n(H-H_{c,n}) \ .
\label{DeltanH}
\end{equation}
Whether the instability field $H_{c,n}$ corresponds to a real phase transition depends on its relation
with the instability fields in other $n$ subsectors including $n=1$.

The instability field $H_{c,n}$ can be related to $H_{c1}$ (\ref{Hc1}) by introducing the energy gap between the lowest
point on the $n$-magnon SIBS  branch $\epsilon_{n}({\bf k}_n)$
and the lowest point in the $n$-magnon continuum. Generally, these two can correspond to different
values of the total momentum $\bf k$. However, for the class of models considered here,
the nearest-neighbor exchange models on Bravais lattices, the two momenta always coincide.
This property allows us to express $H_{c,n}$ via the binding energy at ${\bf k}_n$:
\begin{equation}
E_{B,n} = E_n^l({\bf k}_n)-\epsilon_{n}({\bf k}_n) \ .
\label{EBn}
\end{equation}
The instability  field is, then, given by
\begin{equation}
H_{c,n}  = H_{c1} + \frac{1}{n}\, E_{B,n}  \ .
\label{Hcn}
\end{equation}
Hence, whenever $E_{B,n}>0$, the condensation of multipolar excitations precedes 
the conventional transition into the canted antiferromagnetic phase. Note, that $E_{B,n}$ can become negative,
while $n$-magnon bound states are still  present at other momenta in the Brillouin zone.

The simple Ising expression (\ref{En0}) shows that for fixed $S>1$, the critical field $H_{c,n}$ is an increasing function
of $n$. Therefore,
\begin{equation}
{H}_{c,2S}>{H}_{c,2S-1}> \ldots    
\end{equation}
Hence, the $L$-magnons with the largest value $n=2S$ condense first.
Using a more refined expression (\ref{EnK}) and neglecting  the hopping term, we find  for the transition field of  
longitudinal magnon condensation in the case of $S>1$:
\begin{equation}
H_{cL} = H_{c,2S} = zJS\Bigl(1+  \frac{\alpha\lambda}{2S-1}\Bigr) .
\end{equation}
For large values of $D$, it becomes larger than the one-magnon condensation field (\ref{Hc1}): $H_{cL}> H_{c1}$.
In this regime, the condensation of longitudinal magnons leads to the formation of the high-field multipolar phase
as schematically shown in Fig.~\ref{fig_BEC}. For $S=1$, the two-magnon SIBS are the same as longitudinal magnons.
In this case the hopping amplitude (\ref{tn}) becomes of the same second order as the renormalized bond energy 
(\ref{an}) and both processes contribute equally to the critical field value:
\begin{equation}
H_{cL} = H_{c2} = zJS\bigl(1+  2\alpha\lambda \bigr),\quad S=1 \,.
\end{equation}

\begin{figure*}[t!]
\centerline{   \includegraphics[width = 0.32\linewidth]{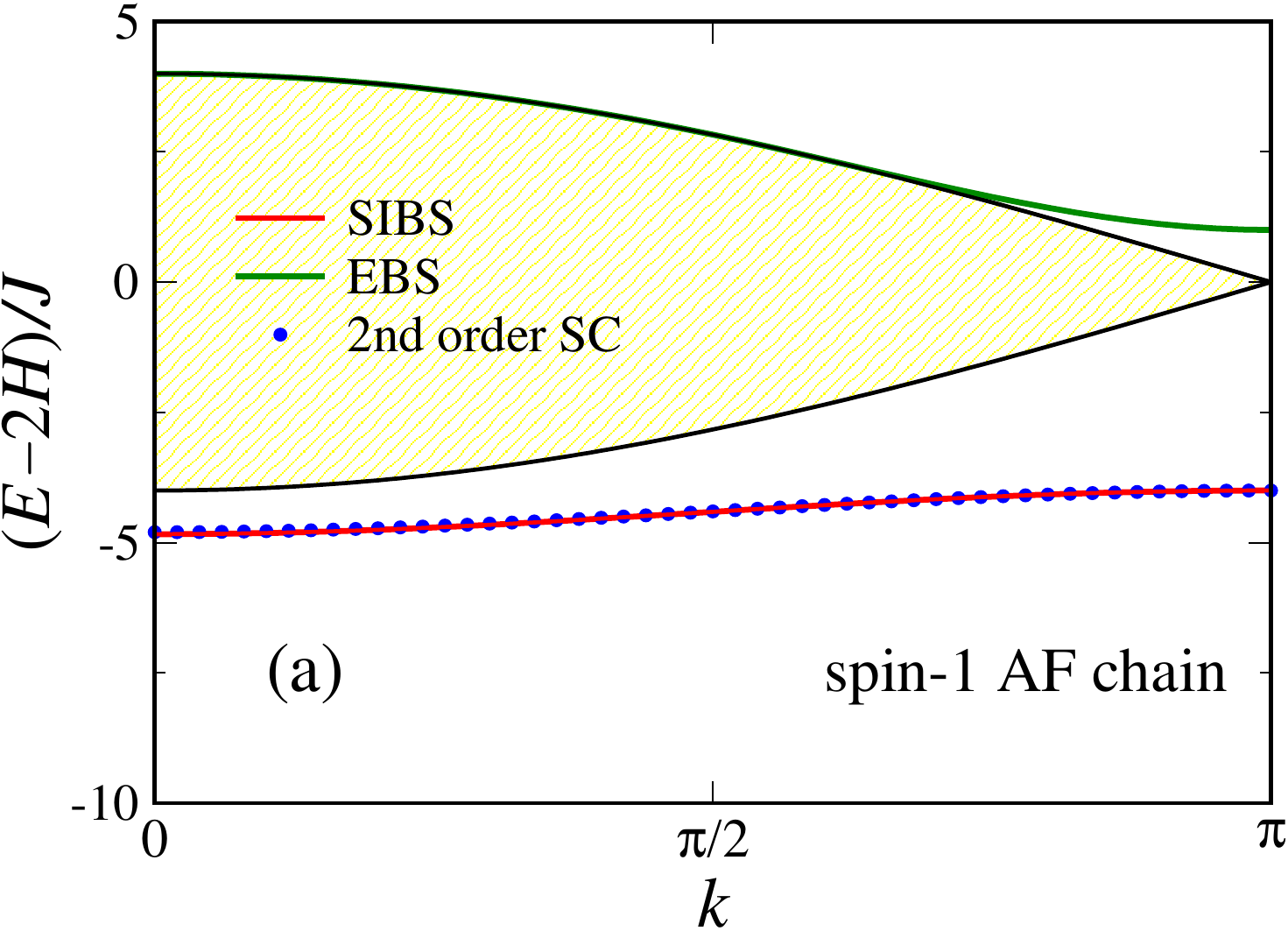}
\hskip 2mm  \includegraphics[width = 0.32\linewidth]{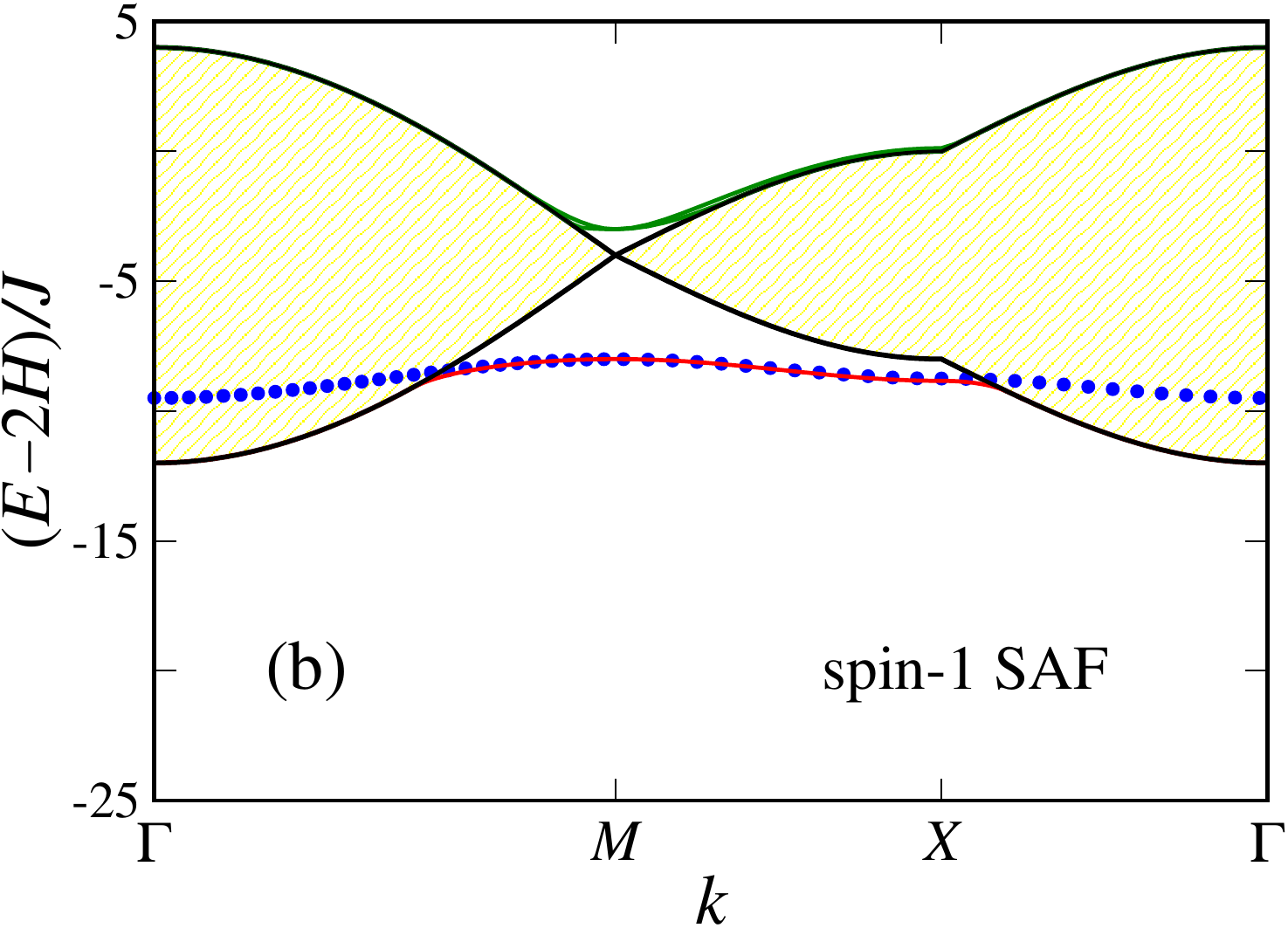}
\hskip 2mm  \includegraphics[width = 0.32\linewidth]{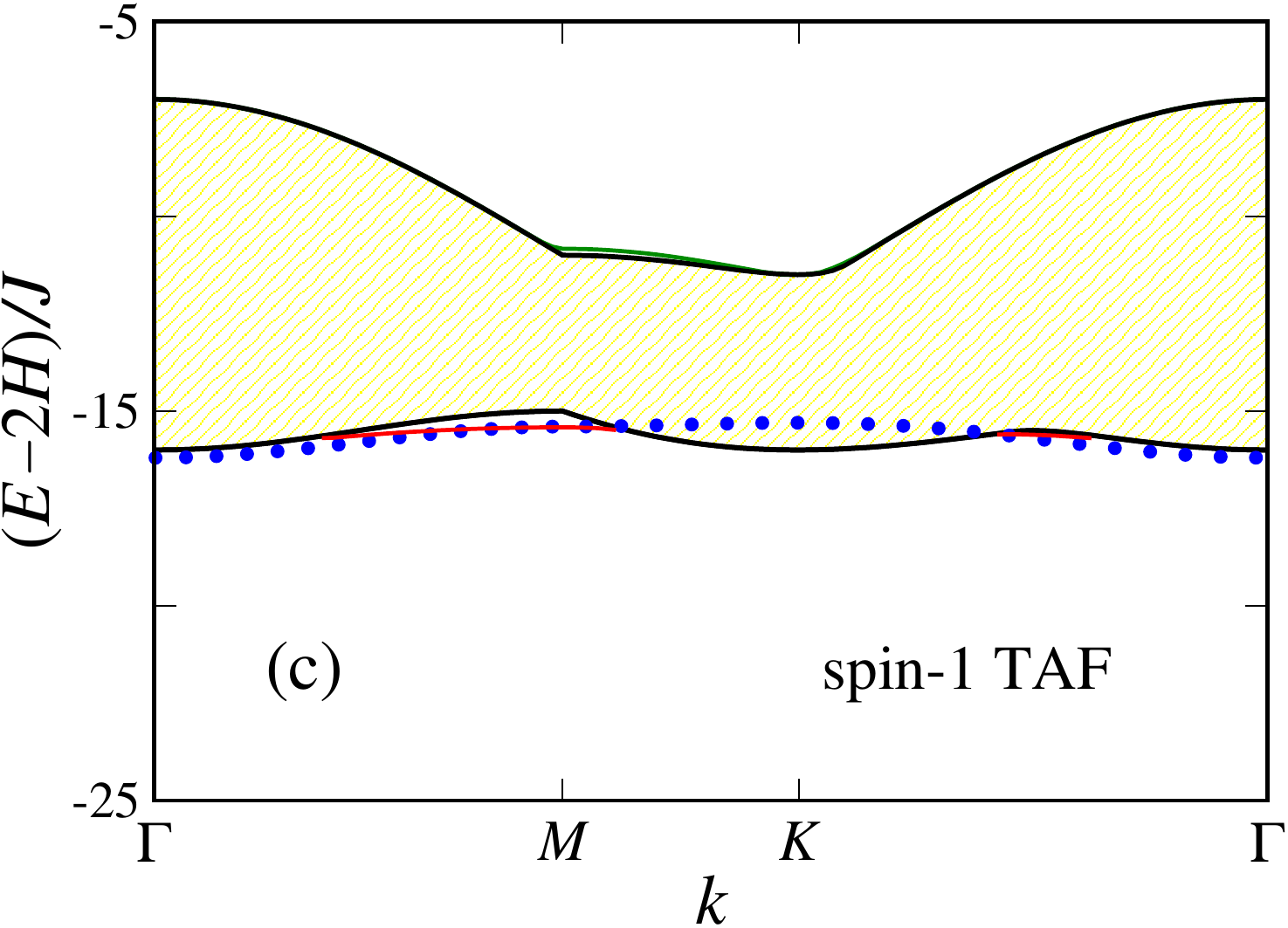}
}
\caption{Two-magnon spectrum in the fully polarized phase of the nearest-neighbor
Heisenberg antiferromagnet for $S=1$ and $D/J = 2$:
(a) one-dimensional  chain, (b) square lattice, and (c) triangular lattice. The lightly shaded region indicates the two-magnon continuum.
The exact SIBS and EBS branches outside the continuum are shown by lines. 
The circles correspond to the second-order strong-coupling (SC) result for SIBS.}
\label{fig_SIBS}
\end{figure*}

Finally, let us briefly discuss higher-order corrections to the multimagnon bound state energy. For the spin-$1$ antiferromagnetic 
chain and the square-lattice antiferromagnet considered in the next section, 
the additional corrections appear only at fourth order of perturbation theory and therefore remain small over 
the relevant range of microscopic parameters. 
By contrast, in the triangular-lattice model, additional third-order processes arise from the elementary plaquettes, 
renormalizing the tight-binding parameters in (\ref{EnK}).
Motivated by the new material  Na$_2$BaNi(PO$_4$)$_2$ discussed in Sec.~\ref{sec:NaBaNiPO}, 
we quote without derivation  the  third-order results for the $S=1$ triangular-lattice antiferromagnet:
\begin{equation}
a_2  =  \frac{(\lambda J)^2}{4D+2J}\, (1\!-\! 4\alpha)\,, \quad 
 t_2  =  \frac{(\lambda J)^2}{4D+2J}\, (1\!-\! 12\alpha)\,.
 \label{atTAF}
\end{equation}

\subsection{High-energy expansion}
\label{sec:HighE}

We now turn to the easy-plane case, $D<0$. Two single spin flips occupying the same lattice site now 
repel each other. 
In the Ising limit, the corresponding local two-magnon state lies above the continuum formed by two spatially
separated magnons. Thus, for sufficiently large $|D|$, these states form an
isolated branch of high-energy quadrupolar excitations with infinite lifetime. Degenerate perturbation theory can
be constructed for such an excited manifold in the same way as for the low-energy SIBS, provided that
it remains well separated from the other states in the same subsector. In what follows, we restrict the
discussion to the repulsively bound state of two magnons. The repulsively bound excitations with $n>2$  
can be treated similarly.

The Ising energy of a local two-magnon state is 
\begin{equation}
\widetilde \varepsilon_{2,0} = 2H - 4(S-1)|D| - 2zJS  
\label{E20AB}
\end{equation}
and exceed the energy of two separated magnons by $2|D|$.
The second-order bond processes generate the same tight-binding dynamics as in the case of
attractive magnon interaction.
The corresponding amplitudes are  given by
\begin{equation}
\widetilde a_2 = \widetilde t_2 =
-\frac{(\lambda J)^2(2S-1)}
{4|D|-2J}\, .
\label{atAB}
\end{equation}
The sign difference relative to the easy-axis expressions 
is due to the sign change in the energy denominator.  Overall,
the coupling to lower-energy  states pushes the repulsively bound branch further upward.
In addition, its minimum is always located at ${\bf k}={\bf Q}$.

\begin{figure*}[t!]
\centerline{   \includegraphics[width = 0.32\linewidth]{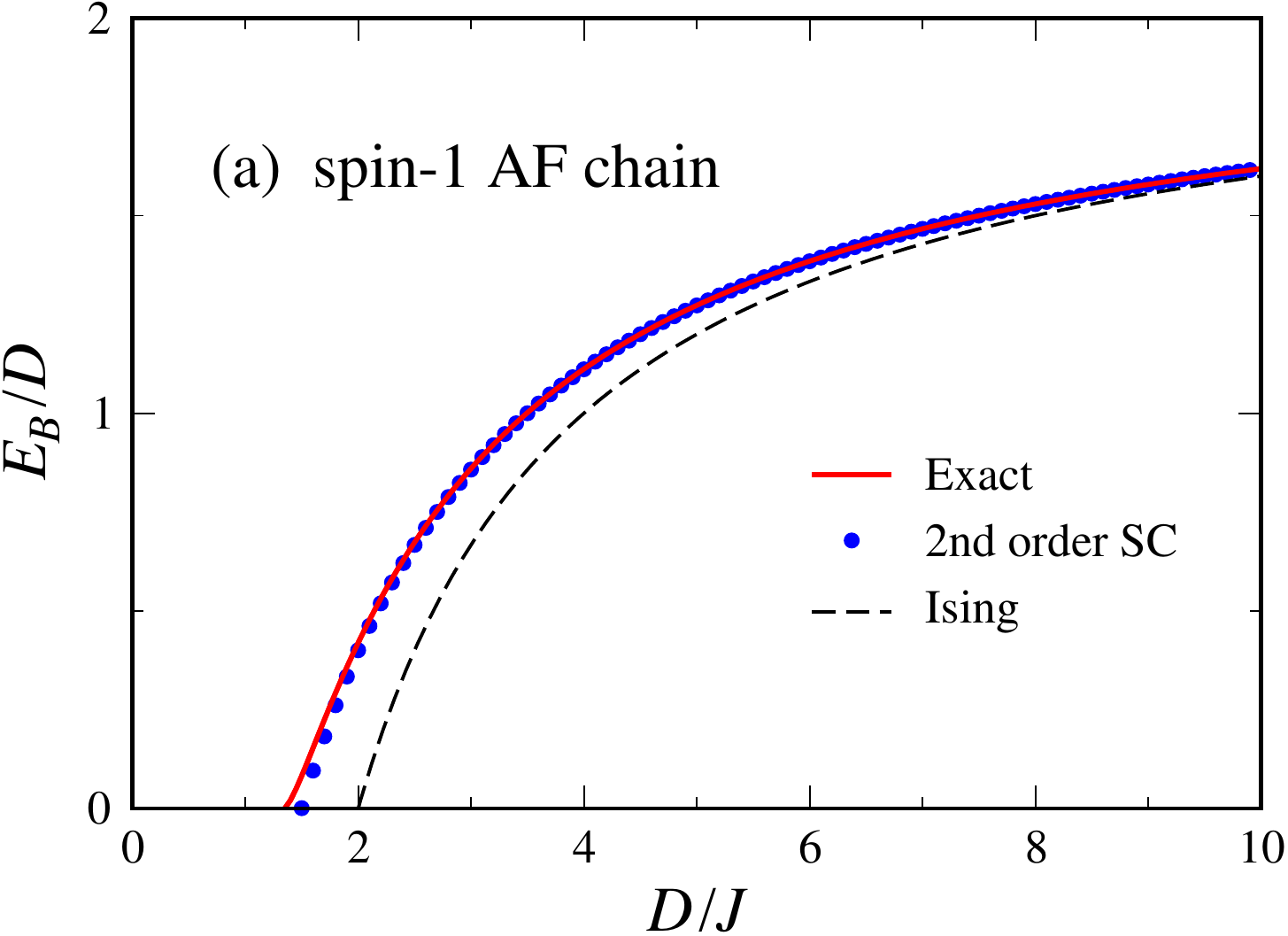}
\hskip 2mm  \includegraphics[width = 0.32\linewidth]{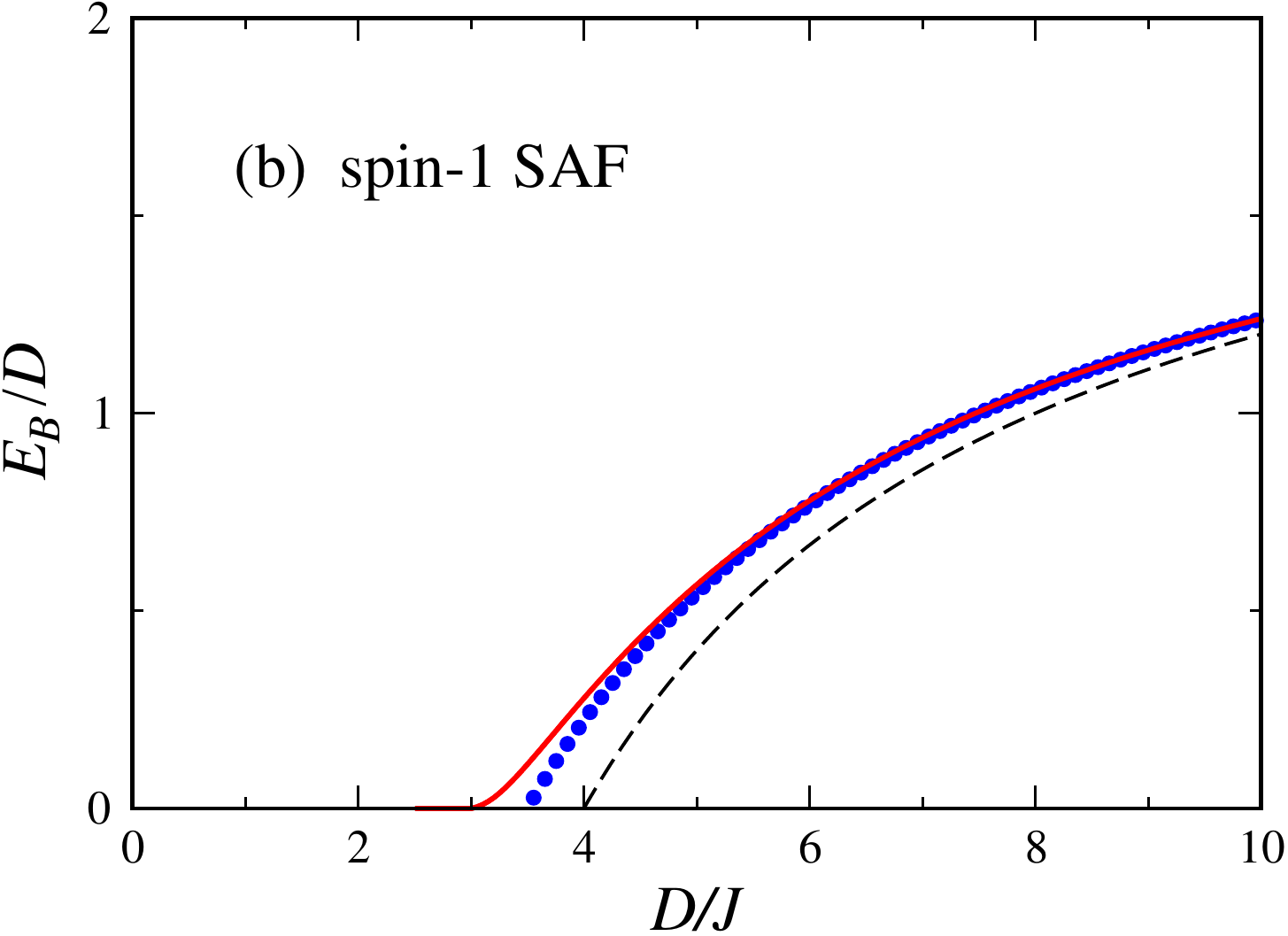}
\hskip 2mm  \includegraphics[width = 0.32\linewidth]{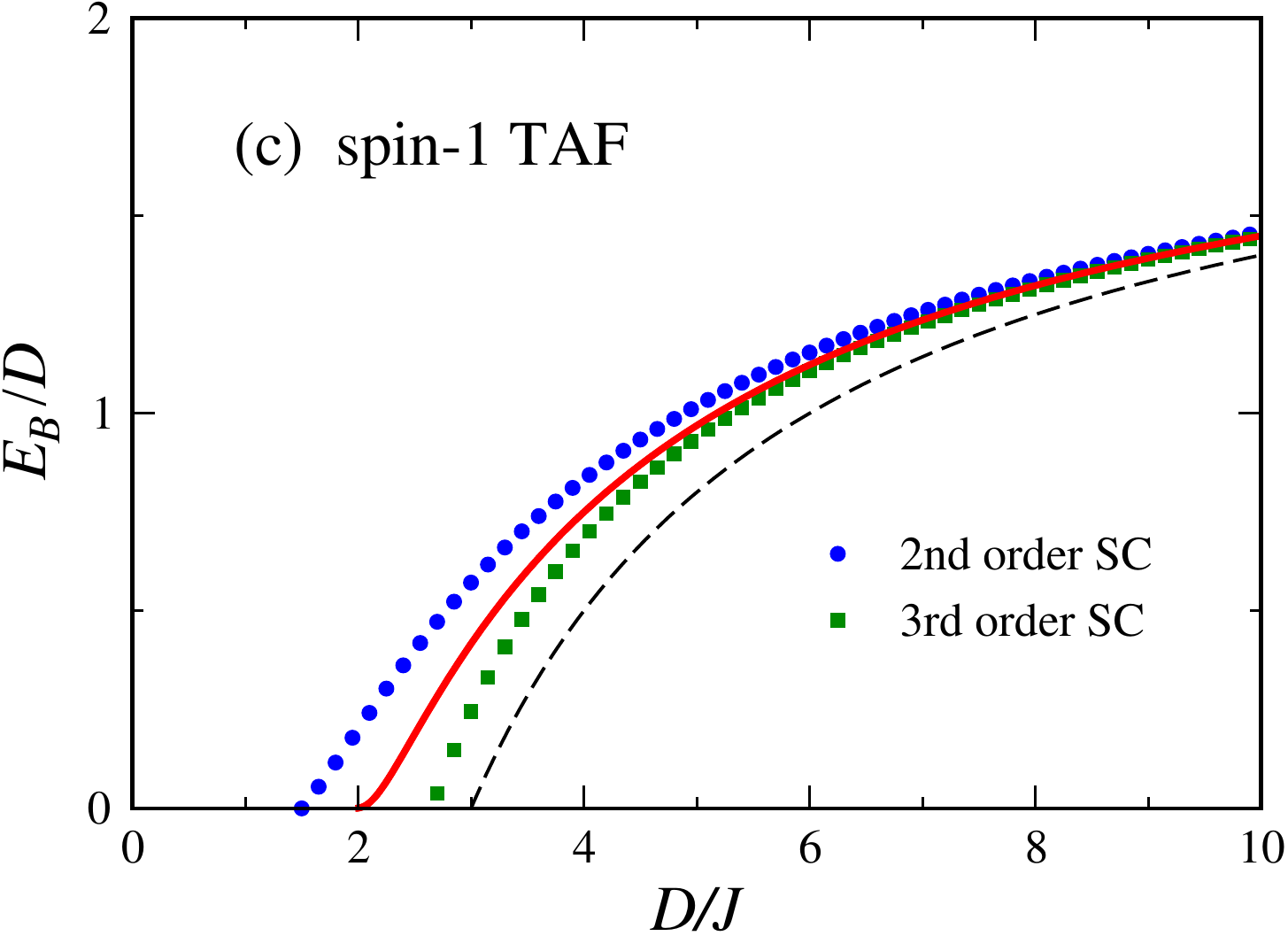}
}
\caption{The ${\bf k}=0$ binding energy of the single-ion bound  magnon states 
in the fully polarized phase of the nearest-neighbor
Heisenberg antiferromagnet with $S=1$  on
(a) one-dimensional  chain,  (b)  square lattice, and (c) triangular lattice.
The exact results are compared with the  strong-coupling expansion and 
    the Ising approximation.}
\label{fig_Eb1}
\end{figure*}

\section{Bound-state spectra and multipolar instabilities}
\label{sec:Num}

We now present the numerical solutions of the exact two-magnon equations of Sec.~\ref{sec:2mag} 
and  compare them with the
approximate theory of Sec.~\ref{sec:Multi}. We further use these results to identify the leading instability 
of the fully polarized phase. For $S=1$, the single-ion bound state is unique and consists of  two magnons 
attracted to the same site. Its dispersion and binding energy can be obtained exactly and compared directly 
with the strong-coupling expansion.  We consider the antiferromagnetic (AF) chain, square-lattice antiferromagnet (SAF), and triangular-lattice antiferromagnet (TAF). We then extend this discussion to $S>1$, 
where several $n$-magnon SIBS can exist and compete
for the leading instability channel as the magnetic field is decreased.

\subsection{Spin-1 antiferromagnets}
\label{sec:S1AFs}
\subsubsection{Two-magnon spectra}
\label{sec:SIBS}

Figure~\ref{fig_SIBS} shows the exact two-magnon spectra for the three $S=1$ antiferromagnets 
with $D/J=2$ and $\lambda=1$.  We use the standard notations  for high-symmetry points in the Brillouin zone:
$\Gamma=(0,0)$, $X=(\pi,0)$, and $M=(\pi,\pi)$ for the square lattice, and 
$\Gamma=(0,0)$, $K=(4\pi/3,0)$, and $M=(\pi,\pi/\sqrt{3})$ for the triangular lattice,  
with the lattice spacing set to unity.
For all three examples, a branch of SIBS is present over at least part of 
the Brillouin zone. As expected, the magnon pairing is strongest in one dimension, 
where the SIBS branch lies outside the continuum for all $k$.
The strong-coupling result (\ref{EnK})  reproduces correctly the SIBS dispersion as long as it appears outside the continuum.
The agreement worsens close to the crossing points, where the two-particle decay singularity forces a tangential 
merging of $\epsilon_2({\bf k})$ with the continuum boundary  \cite{MZH06}.

As discussed in Sec.~\ref{sec:2mag},  for the AF chain and SAF models 
the two magnon continuum  collapses to a single pinch point  at the Brillouin zone boundary ${\bf k}^*$
producing a characteristic bow-tie structure in the energy-momentum plane.
At this momentum,  the bound states remain separated from the continuum and recover the Ising-like level
structure with the exact binding energies $E_B = 2D$ for SIBS and  $E_B = J$ for EBS.
Thus, for nearest-neighbor hypercubic spin models the SIBS branch survives close to ${\bf k}^*$ even when the anisotropy is too weak to keep it outside the continuum in other parts of the Brillouin zone.

No analogous momentum with a protected Ising-like spectrum is present for the triangular-lattice antiferromagnet.
The SIBS branch first splits from the continuum near the $M$ point for $D/J\approx 1.84$ and becomes fully isolated  
from the continuum for $D/J \approx 2.4$.
These two values should be distinguished from the threshold for magnon pairs condensation considered below.
Repulsively bound magnon states are also present in Fig.~\ref{fig_SIBS}, most clearly for the AF chain and progressively less separated from the continuum for the SAF and TAF.

\subsubsection{Pair condensation and spin nematic}
\label{sec:nematic}

The existence of a distinct SIBS branch over a finite part of the Brillouin zone does not by itself imply a spin-nematic instability.
The two-magnon spectrum of the Heisenberg SAF for $D/J=2$, Fig.~\ref{fig_SIBS}(b), provides an example of such a situation.
Condensation of magnon pairs occurs only if the minimum of the SIBS branch 
lies below the global minimum of the two-magnon continuum.
For the three  considered models, the two minima occur at the same total momentum.
Their separation is therefore equal to the binding energy (\ref{EBn}), denoted below simply by $E_B$. 
The corresponding instability field is
\begin{equation}
H_{c2} = H_{c1}+\frac{1}{2}\,E_B \  .
\label{Hc21}
\end{equation}

The evolution of the exact binding energy  is shown in Fig.~\ref{fig_Eb1}.
The critical anisotropy at which the pair instability  overcomes the conventional one-magnon BEC  is
\begin{equation}
(D/J)_c\!=1.35\ ({\rm AF\,chain}),\ \ 
3.0\ ({\rm SAF}),\ \ 
2.0\ ({\rm TAF}).
\label{Dc_spin1}
\end{equation} 
The phase below  $H_{c2}$ is a ferronematic with 
\begin{equation}
\langle S_i^-\rangle=0\quad \textrm{and} \quad \langle(S_i^-)^2\rangle\neq 0 \ .
\end{equation}
The tendency toward bound-state formation is suppressed with increasing dimensionality and therefore requires a larger magnon attraction for magnon-pair condensation to become the dominant instability channel. 
However, this effect is not a simple monotonic function of $z$, as a comparison between the SAF and TAF demonstrates, and can be affected by magnetic frustration.

The critical values (\ref{Dc_spin1}) are obtained from the level crossing and, hence, are equivalent
to a linear approximation in the magnon (pair) density. For our problem, this
approximation is valid for either small or large $D$, where both transitions, at $H_{c1}$ and $H_{c2}$,
are continuous phase transitions. The longitudinal magnons (2-magnon SIBS) have a hard-core repulsion
and are expected to condense continuously for all $D>D_c$. On the other hand, the single-magnon BEC  can become
a first-order transition starting at  certain  $D^*\alt D_c$.
This effect reduces the stability region of the ferronematic phase and shifts the actual threshold  to a somewhat
larger $D/J$. For the $S=1$ Heisenberg SAF, quantum Monte Carlo simulations \cite{Wierschem12} place this threshold at 
$(D/J)_c\approx 3.7$, above the two-particle estimate $(D/J)_c= 3.0$.
No similar results are currently available for the AF chain and TAF models.
Thus, Eq.~(\ref{Dc_spin1}) provides only an estimate of the threshold value for each model, rather than the exact many-body 
boundary.

Finally, let us assess the accuracy of the perturbation approach of Sec.~\ref{sec:Multi}  by comparing successive approximations 
for $E_B$ with the exact two-magnon result in Fig.~\ref{fig_Eb1}.
The Ising approximation is obtained by setting $\varepsilon_{n,0}=0$ in Eq.~(\ref{En0}).
The exact $E_B$  is well approximated by the Ising value for $D/J \agt 5$ . 
The second-order strong-coupling result follows from Eqs.~(\ref{EnK})--(\ref{tn}) with $n=2$ and remains quantitatively 
accurate even for moderate anisotropy, although its accuracy worsens upon approaching the threshold value $(D/J)_c$. 
The deviations are most pronounced for the TAF, where the third-order correction (\ref{atTAF}) produces a noticeable 
shift of the $E_B(D)$ curve toward the exact result. Nevertheless, higher-order terms of the linked-cluster expansion 
are still necessary close to $(D/J)_c$.

\begin{figure}[tb]
\centerline{ \includegraphics[width = 0.8\linewidth]{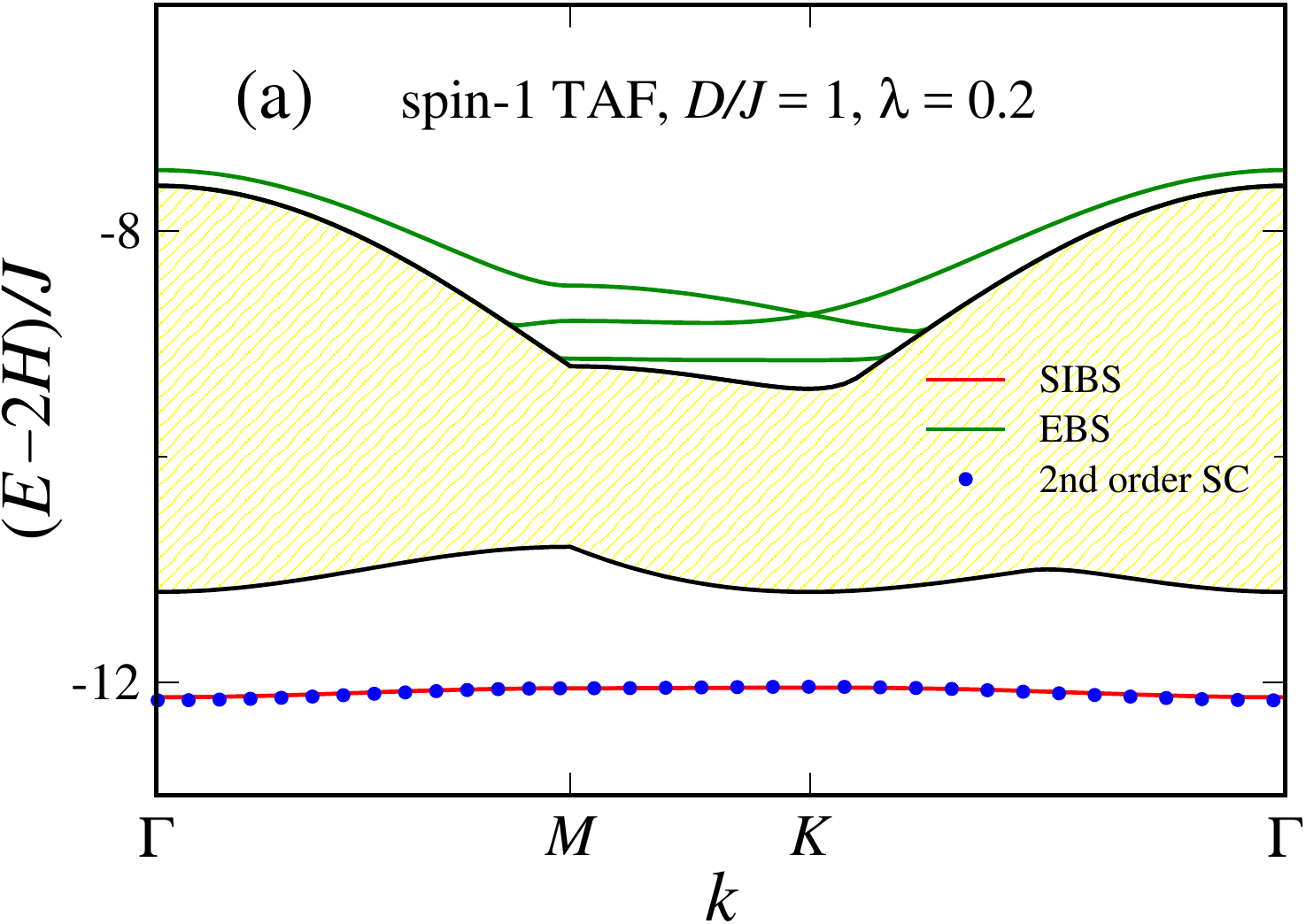}}
\vskip 5mm  
\centerline{ \includegraphics[width = 0.8\linewidth]{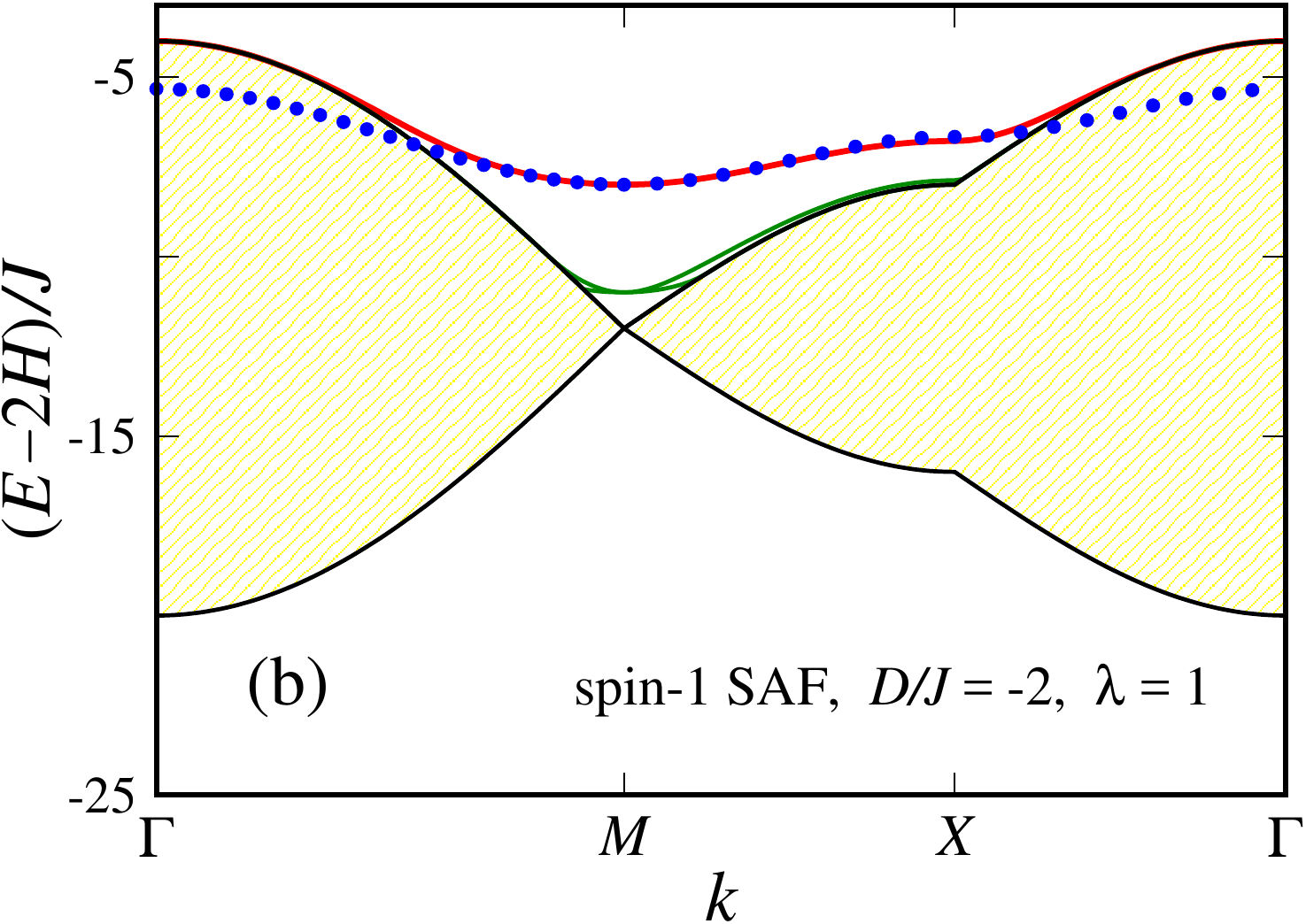}}
\caption{Repulsively bound two-magnon states in the polarized phase of the nearest-neighbor 
$S=1$  antiferromagnet on  (a)
triangular lattice for $D=J$ and $\lambda=0.2$, 
and (b) square lattice for $D/J = -2$ and  $\lambda=1$.
The line and symbol conventions are the same as in Fig.~\ref{fig_SIBS}.
}
\label{fig_EBS}
\end{figure}

\subsubsection{Repulsively bound magnons}
\label{sec:repulsive}

Figure~\ref{fig_EBS} shows two complementary realizations of repulsively bound magnons. The top panel, Fig.~\ref{fig_EBS}(a),
presents the two-magnon spectrum for the TAF with $D=J$ and $\lambda=0.2$. Close to the Ising limit,
there are three distinct exchange-bound branches above the continuum. They provide a direct illustration of
the bond rule of Sec.~\ref{sec:2mag}: the triangular lattice has three independent bond directions 
and, hence, supports three EBS branches despite being two dimensional. 
As $\lambda$ increases, the magnon continuum broadens, forcing the lowest EBS branch to disappear 
at the critical value $\lambda_{c1}\approx 0.5$. The two remaining branches coexist up to  
$\lambda_{c2}\approx 0.83$. For larger $\lambda$, there is at most one bound state
 per $\bf k$ above the continuum.  Note that these  $\lambda_c$ values have been determined for $S=1$ 
and can change substantially for other values of spin.

Figure \ref{fig_EBS}(b) shows the two-magnon spectrum for the Heisenberg SAF with $D/J=-2$. 
It illustrates the repulsively bound magnon state pushed out of the continuum by
the strong easy-plane anisotropy.   
The high-energy strong-coupling expansion of Sec.~\ref{sec:Multi} follows closely the exact dispersion,
apart from the  vicinity of the crossing points into the continuum, where processes beyond second order
become important. Note that
a pair of EBS branches exist near the $M$ point. The Ising values for the binding energies
reemerge here as well. To sum up, the same two-magnon formalism describes discrete states below 
or above the continuum, depending on whether the underlying magnon interaction is attractive or repulsive.

\subsection{Large-$S$ antiferromagnets}
\label{sec:higherS}

For $S>1$, the single-ion anisotropy generates a hierarchy of local $n$-magnon bound states
with $n=2,\ldots,2S$. Their energies can be computed using the strong-coupling expansion of Sec.~\ref{sec:Multi}.
For even $n$ the band minimum of $\epsilon_{n}({\bf k})$ is located at ${\bf k}=0$,
whereas for odd $n$ it occurs at the antiferromagnetic ordering wavevector ${\bf Q}$.
For SAF considered below, the $n$-magnon continuum has the
global minimum at the same total momentum as $\epsilon_{n}({\bf k})$,
hence positive binding energy $E_{B,n}({\bf k}_n)$ directly
implies $H_{c,n}>H_{c1}$.

\begin{figure}[tb]
\centerline{\includegraphics[width=0.8\linewidth]{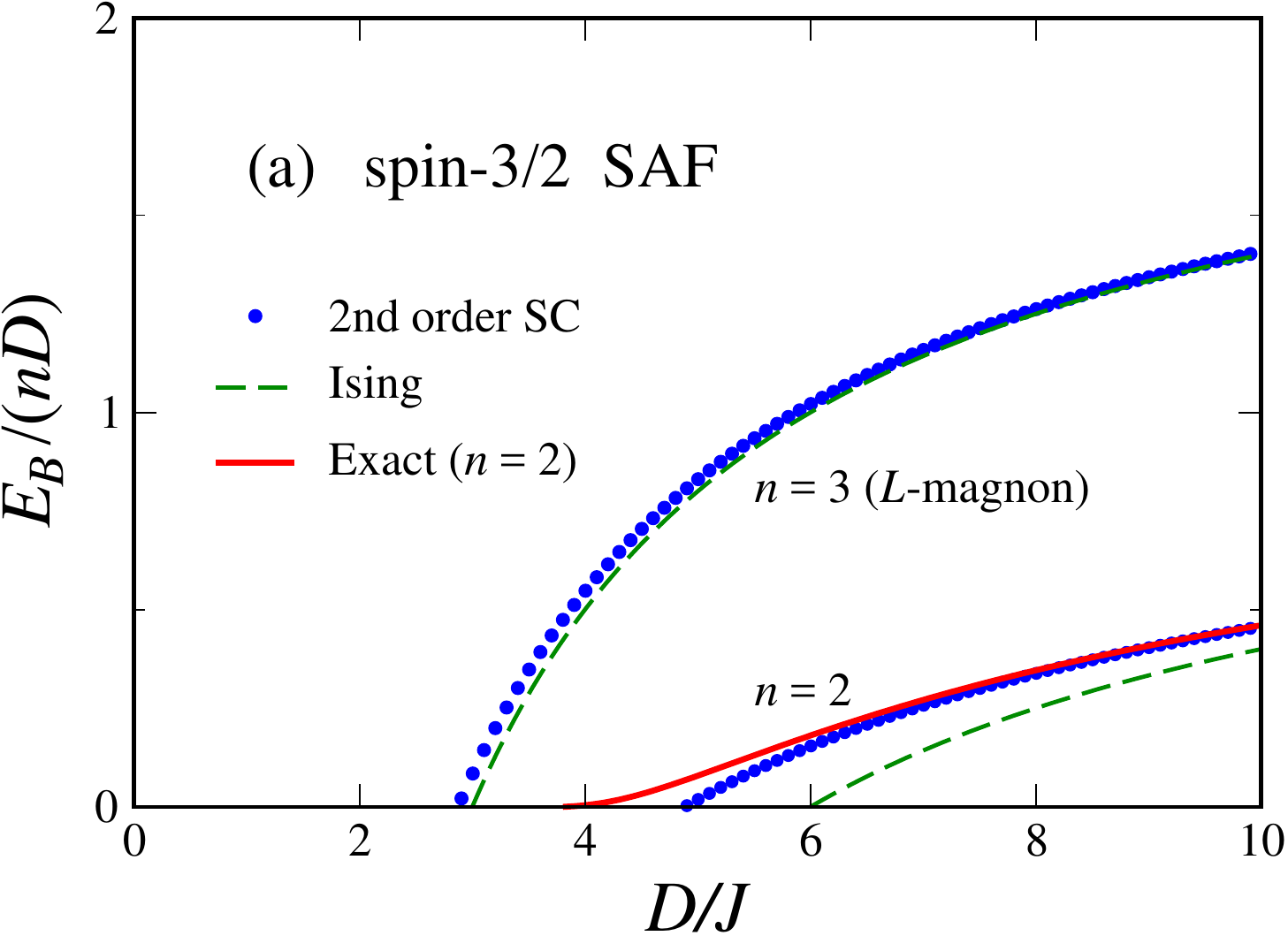}}
\vskip 5mm
\centerline{\includegraphics[width=0.8\linewidth]{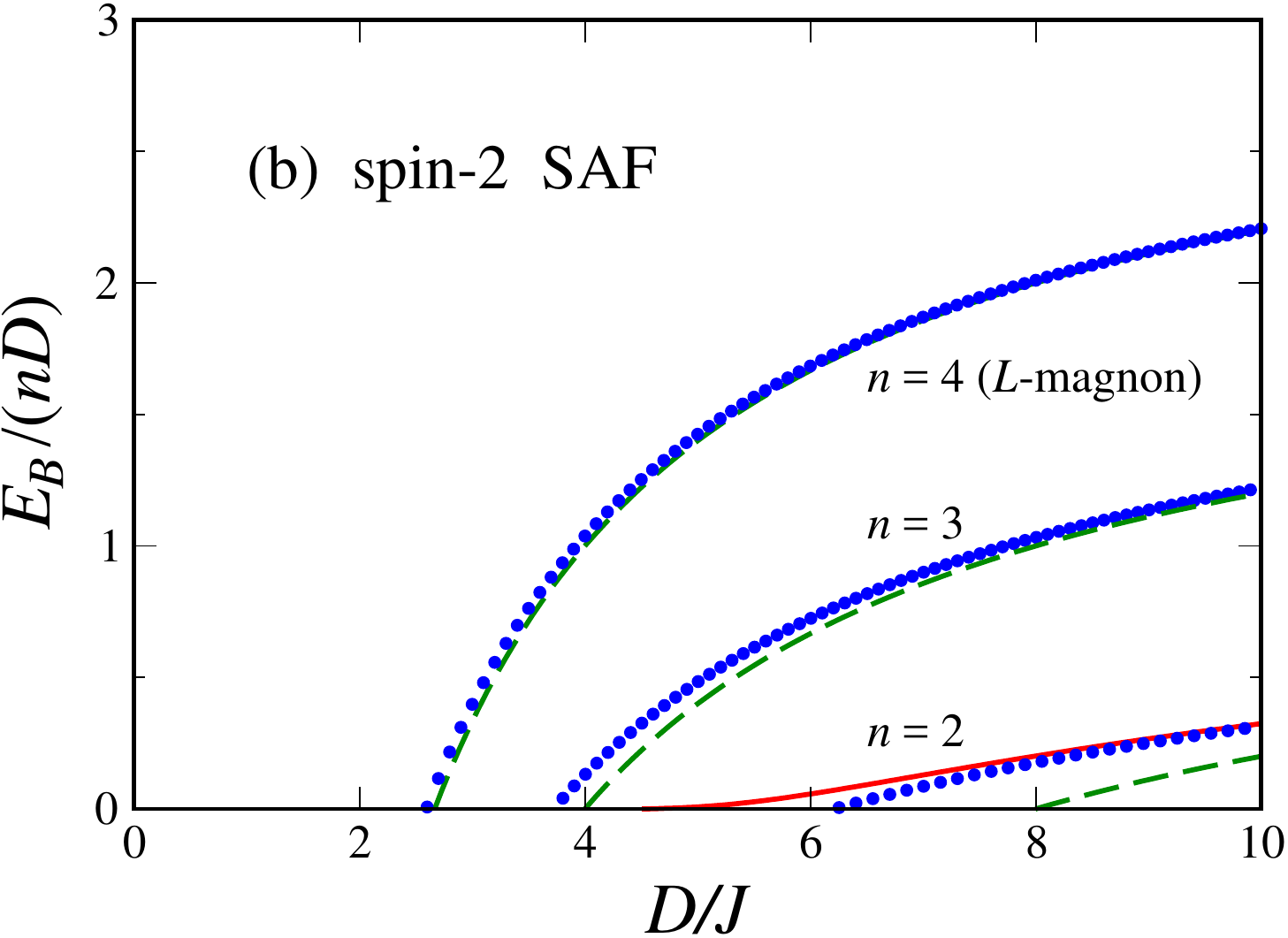}}
\caption{Binding energies of the $n$-magnon SIBS in fully polarized phase of the Heisenberg antiferromagnet
($\lambda =1$) on a square lattice for (a) $S=3/2$  and  (b) $S=2$.
For $n=2$, the strong-coupling result is compared with the exact two-magnon solution.
}
 \label{fig_EbS}
\end{figure}

Figure~\ref{fig_EbS} shows the anisotropy dependence of the
binding energies for the allowed $n$-magnon SIBS  in the Heisenberg  square-lattice antiferromagnets 
for two values of spin $S=3/2$ and $S=2$.
Each $n$-magnon bound state becomes stable above its own threshold value of $D/J$, while the largest value of $E_{B,n}/n$ determines the highest instability field $H_{c,n}$.
For $S>1$, the longitudinal magnon with $n=2S$ determines the leading multimagnon instability 
above the anisotropy threshold, see Fig.~\ref{fig_BEC}.
The resulting multipolar phase emerges below $H_{cL}=H_{c,2S}$ and is characterized by the multipole order parameter
\begin{equation}
\langle (S_i^-)^{2S} \rangle \neq 0 \ .
\label{Multipole}
\end{equation}
Such states provide a natural higher-rank generalization of the spin-nematic phase realized for $S=1$.
The long-range order is carried by composite spin operators rather than by the conventional transverse dipole moments, 
making the multipolar states (\ref{Multipole}) a particularly interesting class of quantum magnetic phases for $S>1$.

Figure~\ref{fig_EbS} also provides a direct test of the strong-coupling approximation.
For the two-magnon SIBS, the binding energy can be obtained independently from the exact Schr\"odinger equation.
Its comparison with the second-order result demonstrates good quantitative accuracy of the strong-coupling expansion except in the immediate vicinity of the bound-state threshold.
For increasing $n$, the SIBS become progressively more localized and their bandwidth decreases, so that their energies approach more closely the corresponding Ising-limit behavior.

To sum up, the pair-condensation mechanism discussed above for spin-$1$ systems represents the first member of a broader hierarchy of field-induced multipolar instabilities generated by the easy-axis single-ion anisotropy, see Fig.~\ref{fig_BEC}.

\section{Experimental Applications}
\label{sec:Exp}

We now apply the developed theory to two recently studied materials that provide
experimental examples of the single-ion bound states of magnons formed by attractive and repulsive interactions. 
The easy-axis triangular-lattice antiferromagnet Na$_2$BaNi(PO$_4$)$_2$ realizes Bose-Einstein 
condensation of  bound states in a magnetic field, 
whereas the easy-plane square-lattice antiferromagnet Sr$_2$CoGe$_2$O$_7$ exhibits repulsively bound 
magnon pairs above the continuum.

\subsection{Bound-magnon condensation in Na$_2$BaNi(PO$_4$)$_2$}
\label{sec:NaBaNiPO}

The newly synthesized spin-1 antiferromagnet Na$_2$BaNi(PO$_4$)$_2$, with magnetic ions forming an ideal triangular lattice \cite{Li21}, exhibits an unusual field-induced transition driven by the Bose-Einstein condensation of bound magnon pairs 
\cite{Sheng25a,Sheng25b}. This observation provides compelling experimental evidence  for the long-sought quantum spin-nematic phase \cite{Blume69,Andreev84,Penc11}.  
A relatively small saturation field of $B_s\approx1.8$~T ($B\parallel c$) enables accurate determination of the microscopic constants via inelastic neutron-scattering measurements in the fully polarized phase:
\begin{equation}
J = 0.036\ ,\quad \lambda J = 0.032\ , \quad 
D =  0.127
 \label{NaBaNiPO}
\end{equation}
(all in meV) \cite{Sheng25a}.
The pronounced easy-axis anisotropy with $D/J \approx 3.51$ and $\lambda \approx 0.885$ fully agrees with the 
 magnetization measurements \cite{Li21}.
 
The one-magnon condensation field (\ref{Hc1}) for $S=1$ TAF is
\begin{equation} 
g\mu_B B_{c1} = 3J(2+\lambda) - D \,.
\end{equation}
Using $g_c=2.24$ together with the parameters (\ref{NaBaNiPO}), we obtain
$B_{c1} = 1.434$~T, which is significantly below the observed transition at
$B_s \approx1.8$~T. Hence, the high-field transition in the fully polarized phase cannot be 
assigned to the one-magnon instability.

The exact solution of the two-magnon Schr\"odinger equation with the parameters (\ref{NaBaNiPO})
reveals a branch of single-ion bound magnon states below the continuum.
It is the condensation of quasiparticles from this branch which determines the instability of 
the fully polarized phase upon decreasing the applied field. The large easy-axis anisotropy, corresponding to a small 
Ising parameter  $\alpha\approx0.055$,
places the system well within the strong-coupling regime. The exact dispersion of SIBS is 
very accurately reproduced by the third-order perturbation result (\ref{atTAF}). 
Setting the excitation gap to zero gives
the following expression for the two-magnon condensation field:
\begin{equation}
g\mu_B B_{c2} = 6J\bigl(1+2\lambda\alpha-16\lambda\alpha^2\bigr) .
\label{Hc2}
\end{equation}
For the experimental set of parameters, one finds 
$B_{c2} \approx 1.77$~T in good agreement with the measured $B_s$.

\subsection{Repulsively bound magnons in Sr$_2$CoGe$_2$O$_7$}
\label{sec:SrCoGeO}

Sr$_2$CoGe$_2$O$_7$ provides a complementary example of spin-quadrupolar excitations generated by repulsive magnon interactions. The magnetic Co$^{2+}$ ions in this material are described by effective spins $S=3/2$.
They are arranged on a square lattice and interact via anisotropic nearest-neighbor exchange with a significant  easy-plane anisotropy and negligible interplane coupling. The microscopic parameters determined from the magnetization and ESR measurements  \cite{Akaki17} are given by
\begin{equation}
J=45\pm1.1,\quad
\lambda J=49.1\pm0.8,\quad
D=-49.7\pm4.2
\label{SrCoGeO}
\end{equation}  
(all in GHz).
High-field ESR measurements in the fully polarized phase above 
$B_s\approx 21.5$~T ($B\parallel c$)
resolve several one-magnon modes with an effective $g$-factor $g_c\approx 2.23$ and
 an additional excitation whose field dependence is described by a twice larger $g_c^*\approx  4.42$  
\cite{Akaki17}. This extra mode was identified as a repulsively bound  state of two magnons lying above the two-magnon continuum.  
  
For the microscopic parameters (\ref{SrCoGeO}), the exact solution of the two-magnon Schr\"odinger equation 
finds no bound state at zero momentum. A repulsively bound single-ion state does, however, exist near ${\bf k}=(\pi,\pi)$, where the protected Ising  spectrum is realized, see Sec.~\ref{sec:2mag}. Excitations at ${\bf k}=(\pi,\pi)$ 
become ESR active in Sr$_2$CoGe$_2$O$_7$ because the alternating rotation of the oxygen tetrahedra 
doubles the crystallographic unit cell in this material and folds the $(\pi,\pi)$  momentum to the center of the reduced Brillouin zone.  
  
At the saturation field $B_s$, the gap of the lowest one-magnon mode  $\nu_{1a}$ vanishes,
whereas the second one-magnon mode $\nu_{1b}$ and the anomalous two-magnon mode $\nu_2$
have the following gaps \cite{Akaki17}
\begin{equation}
\Delta_{1b}\approx580~{\rm GHz},\quad
\Delta_2\approx690~{\rm GHz}\,.
\end{equation}
The gapless mode $\nu_{1a}$ corresponds to a magnon with ${\bf k}=(\pi,\pi)$ which condenses
at $B_s$. The  mode $\nu_{1b}$ is associated with the ${\bf k}=0$ magnon. 
For the same field, the pinch point in the two-magnon continuum corresponding to the total momentum ${\bf k}=(\pi,\pi)$ appears precisely at the energy $\Delta_{1b}$. The exact Ising-spectrum protection discussed in Sec.~\ref{sec:ES} then implies that the repulsively bound single-ion state lies above this continuum with
\begin{equation}
\Delta_2-\Delta_{1b}=2|D| \,.
\label{SrCoGeO_bind}
\end{equation}
For the reported anisotropy constant (\ref{SrCoGeO}), the right-hand side is approximately
$99.5\pm 8$~GHz, in a reasonable agreement with the measured gap difference
$\Delta_2-\Delta_{1b}\approx110$~GHz.

Thus, the observed repulsively bound magnon state in Sr$_2$CoGe$2$O$7$ provides a direct realization 
of the protected Ising spectrum at the pinch point of the two-magnon
continuum in a square-lattice antiferromagnet. The same argument predicts an exchange bound state at
\begin{equation}
\Delta_{2b} = \Delta_{1b} + J\approx 625~{\rm GHz}\,.
\end{equation}
This excitation has not been resolved experimentally so far, most likely because its ESR spectral weight is strongly suppressed by the selection rules relevant to this material \cite{Akaki17}.

\section{Conclusions}
\label{sec:Sum}

We have developed a unified theoretical framework for multimagnon bound states in large-$S$ 
antiferromagnets with strong easy-axis anisotropy. Starting from the exact two-magnon Schr\"odinger 
equation, we describe both single-ion and exchange bound states generated by attractive or repulsive 
magnon interactions.  As an additional mathematical result, we obtain rigorous Gershgorin bounds 
on their binding energies. For higher magnon sectors, a strong-coupling expansion yields 
the dispersions and condensation fields of $n$-magnon single-ion bound states with $2\leq n\leq2S$. 
The obtained picture establishes a hierarchy of field-induced instabilities.
For $S=1$, sufficiently strong easy-axis anisotropy causes two-magnon single-ion bound states to condense 
before the one-magnon mode, producing a quantum spin-nematic phase.
For $S>1$, the same mechanism extends naturally to longitudinal $2S$-magnon states and higher multipolar order.

An important direction for future studies is investigating the role of interactions between composite quasiparticles
in the formation and stability of the multipolar order. For the spin-1/2 $J_1$--$J_2$ Heisenberg SAF, where
Bose-Einstein condensation of the exchange-bound magnon pairs occurs near the saturation field \cite{Shannon06},
the canted antiferromagnetic state is stabilized by interactions beyond its classical boundary up to $H^*>H_{c1}$
\cite{Jiang23}. This confines the spin-nematic phase to a rather narrow interval of applied magnetic fields $H^*<H<H_{c2}$,
contrary to the naive picture shown in  Fig.~\ref{fig_BEC}.
The condensation of single-ion bound states in magnetic systems with $S\geq 1$ may avoid this limitation.
In particular, the finite-dimensional Hilbert space excludes two longitudinal magnons on the same site. This restriction 
gives rise to a hard-core repulsion between the quasiparticles, which in turn supports a continuous BEC transition 
into a multipolar superfluid phase. At the same time, conventional 
dipolar ordering becomes energetically disfavored by the large anisotropy to exchange ratio $D/J$. 

Counterintuitively, large-$S$ easy-axis antiferromagnets, which are usually regarded as nearly classical systems, 
may provide a suitable experimental platform for realizing exotic quantum phases. 
The spin-1 triangular antiferromagnet Na$_2$BaNi(PO$_4$)$_2$ is one such example  \cite{Sheng25a,Sheng25b}. 
The general nature of these new exotic phases---multipolar superfluids, supersolids, or Wigner crystals, associated with fractional magnetization plateaus,---will depend on the residual interactions between the composite quasiparticles.
Studying these interactions and determining the many-body phase diagrams presents 
a major challenge for future theoretical work.
We refer interested readers to several previous studies in this area \cite{Damle06,Sen09} as well as a recent preprint 
\cite{Huang26}.

Finally, we note that an additional level of hierarchy in the excitation spectra may arise when the 
multimagnon single-ion quasiparticles themselves form bound states via exchange attraction. For example,
in the spin-1 antiferromagnet FeI$_2$, the observed four- and six-magnon excitations  were interpreted 
as, respectively, bound states of two and three full spin reversals  (longitudinal magnons) \cite{Legros21}.
 A different possibility was considered in theoretical works on the easy-axis $S=1$ ferromagnetic chain  
 \cite{Sharma22,Wang26}, where a distinct  three-magnon excitation
forms due to the binding of one longitudinal magnon with a conventional transverse magnon. Thus, 
single-ion and exchange binding mechanisms can act successively, generating hybrid excitations built from 
already composite magnetic quasiparticles.

\acknowledgments
We are grateful to A.~L.~Chernyshev, M.~Mourigal, M.~Orlita,  A.~I.~Smirnov, H.~Tsunetsugu, and T. Ziman for
stimulating discussions  and to P. Kumar for help with numerical computations.

\bibliography{SIBS}

\end{document}